\documentclass[a4paper,11pt]{article}

\usepackage{graphicx}
\usepackage{amsthm,amsmath,amssymb,amsfonts}
\usepackage{mathtools}
\usepackage{xspace}
\usepackage{color}
\usepackage{xcolor}
\usepackage{thmtools}
\usepackage{hyperref}
\usepackage{cleveref}
\usepackage{comment}
\usepackage{framed}
\usepackage{qcircuit}
\usepackage{fullpage}

\mathchardef\mhyphen="2D 

\newcommand{\cqs}{{\rm cq}\mhyphen\Sigma_2}

\newcommand{\HttI}{{H_t^I}}

\newcommand{\HttiX}{H_{t}^{iX}}

\newcommand{\HttV}{{H_t^V}}

\newcommand\newmathabbrev[2]{\newcommand{#1}{\ensuremath{#2}\xspace}}

\newcommand\cfont\mathrm
\newmathabbrev\p{\cfont{P}}
\newmathabbrev{\N}{\mathbb N}
\newmathabbrev\NP{\cfont{NP}}
\newmathabbrev\TFNP{\cfont{TFNP}}
\newmathabbrev\MHS{\cfont{MHS}}
\newmathabbrev\DTIME{\cfont{DTIME}}
\newmathabbrev\tSAT{3\cfont{\mhyphen{}SAT}}
\newmathabbrev\MA{\cfont{MA}}
\newmathabbrev\MAo{\cfont{MA}_1}
\newmathabbrev\AM{\cfont{AM}}
\newmathabbrev\NPDAG{\cfont{NP\mhyphen{}DAG}}
\newmathabbrev\QMADAG{\cfont{QMA\mhyphen{}DAG}}
\newmathabbrev\yes{\mathrm{yes}}
\newmathabbrev\no{\mathrm{no}}
\newmathabbrev\US{\cfont{US}}
\newmathabbrev\FP{\cfont{FP}}
\newmathabbrev\PP{\cfont{PP}}
\newmathabbrev\CeP{\cfont{C_=P}}
\newmathabbrev\coCeP{\cfont{coC_=P}}
\newmathabbrev\PH{\cfont{PH}}
\newmathabbrev\SAT{\cfont{SAT}}
\newmathabbrev\QSAT{\cfont{QSAT}}
\newmathabbrev\hQSAT{\mhyphen\QSAT}
\newmathabbrev\SPP{\cfont{SPP}}
\newmathabbrev\GapP{\cfont{GapP}}
\newmathabbrev\BQP{\cfont{BQP}}
\newmathabbrev\QP{\cfont{QP}}
\newmathabbrev\StoqMA{\cfont{StoqMA}}
\newmathabbrev\coNP{\cfont{coNP}}
\newmathabbrev\AzPP{\cfont{A_0PP}}
\newmathabbrev\QMA{\cfont{QMA}}
\newmathabbrev\QMAo{\cfont{QMA}_1}
\newmathabbrev\cloneQMA{\cfont{clonableQMA}}
\newmathabbrev\QSigma{\cfont{Q}\Sigma}
\newmathabbrev\coQMA{\cfont{coQMA}}
\newmathabbrev\BPP{\cfont{BPP}}
\newmathabbrev\QCMA{\cfont{QCMA}}
\newmathabbrev\pNPlog{\p^{\NP[\log]}}
\newmathabbrev\pNP{\p^{\NP}}
\newmathabbrev\pNPtwo{\p^{\NP[2]}}
\newmathabbrev\pNPone{\p^{\NP[1]}}
\newmathabbrev\pParSAT{\p^{||\SAT}}
\newmathabbrev\pQMApar{\p^{||\QMA}}
\newmathabbrev\pCpar{\p^{||\C}}
\newmathabbrev\pStoqMApar{\p^{||\StoqMA}}
\newmathabbrev\pQMAlog{\p^{\QMA[\log]}}
\newmathabbrev\pClog{\p^{\textup{C}[\log]}}
\newmathabbrev\pC{\p^{\textup{C}}}
\newmathabbrev\QMASPACE{\cfont{QMASPACE}}
\newmathabbrev\pQMAtlog{\p^{\QMA(2)[\log]}}
\newmathabbrev\pStoqMAlog{\p^{\StoqMA[\log]}}
\newmathabbrev\pQMApt{\p^{\Vert\QMA(2)}}
\newmathabbrev\pQMA{\p^{\QMA}}
\newmathabbrev\SharpP{\cfont{\#P}}
\newmathabbrev\pSharP{\p^{\SharpP[1]}}
\newmathabbrev\PromisePP{\cfont{PromisePP}}
\newmathabbrev\lett{\le_\mathrm{tt}}
\newmathabbrev\YES{\mathsf{YES}}
\newmathabbrev\NO{\mathsf{NO}}
\newmathabbrev\PSPACE{\cfont{PSPACE}}
\newmathabbrev\IP{\cfont{IP}}
\newmathabbrev\POLY{\cfont{POLY}}
\newmathabbrev\DAG{\cfont{DAG}}
\newmathabbrev\StoqMADAG{\StoqMA\mhyphen\cfont{DAG}}
\newmathabbrev\CDAG{C\mhyphen\cfont{DAG}}
\newmathabbrev\CDAGf{C\mhyphen\cfont{DAG}_f}
\newmathabbrev\CDAGs{C\mhyphen\cfont{DAG}_s}
\newmathabbrev\CDAGd{C\mhyphen\cfont{DAG}_{d}}
\newmathabbrev\CDAGo{C\mhyphen\cfont{DAG}_1}
\newmathabbrev\LOGS{\cfont{LOGS}}
\newmathabbrev\TAUT{\cfont{TAUTOLOGY}}
\newmathabbrev\SBQP{\cfont{SBQP}}
\newmathabbrev\SBP{\cfont{SBP}}
\newmathabbrev\Fc{F_\coNP}
\newmathabbrev\Fa{F_\AzPP}
\newmathabbrev\GSCON{\cfont{GSCON}}
\newmathabbrev\GSCONexp{\GSCON_\cfont{exp}}
\newmathabbrev\QMAexp{\QMA_\cfont{exp}}
\newmathabbrev\UQMA{\cfont{UQMA}}
\newmathabbrev\R{\mathbb R}
\newmathabbrev\Trees{\cfont{TREES}}
\newmathabbrev\apxsim{\cfont{APX\mhyphen{}SIM}}
\newmathabbrev\AWPP{\cfont{AWPP}}
\newmathabbrev\X{\mathcal{X}}
\newmathabbrev\Y{\mathcal{Y}}

\newmathabbrev\Z{\mathcal{Z}}
\newmathabbrev\ZZ{\mathbb{Z}}
\newmathabbrev\Hprop{H_\mathrm{prop}}
\newmathabbrev\Hin{H_\mathrm{in}}
\newmathabbrev\Piin{\Pi_\mathrm{in}}
\newmathabbrev\Hout{H_\mathrm{out}}
\newmathabbrev\Piout{\Pi_\mathrm{out}}
\newmathabbrev\Hstab{H_\mathrm{stab}}
\newmathabbrev\Lext{\L_\mathrm{ext}}
\newmathabbrev\BTWNP{\cfont{BTW}(\NP)}
\newmathabbrev\BSN{\cfont{BSN}}
\newmathabbrev\SN{\cfont{SN}}
\newmathabbrev\BD{\cfont{BD}}
\newmathabbrev\HYPERTREE{\cfont{NP\mhyphen{}HYPERTREE}}
\newmathabbrev\Hext{H_\mathrm{ext}}
\newmathabbrev\Hpropt{\tilde{H}_\mathrm{prop}}
\newmathabbrev\Hint{\tilde{H}_\mathrm{in}}
\newmathabbrev\Houtt{\tilde H_\mathrm{out}}
\newmathabbrev\EXP{\cfont{EXP}}
\newmathabbrev\A{\mathcal{A}}
\newmathabbrev\U{\mathcal{U}}

\renewcommand\L{\mathcal{L}}

\newmathabbrev\DAGSSAT{\DAGS(\SAT)}
\newmathabbrev\DAGS{\mathrm{DAGS}}
\newmathabbrev\DAGSNP{\DAGS(\NP)}

\newmathabbrev\AND{\cfont{AND}}

\newmathabbrev\STCONN{{S,T}\cfont{\mhyphen{}CONN}}
\newmathabbrev\CNF{\cfont{CNF}}
\newmathabbrev\NEXP{\cfont{NEXP}}
\newmathabbrev\NPSPACE{\cfont{NPSPACE}}
\newmathabbrev\QCMASPACE{\cfont{QCMASPACE}}
\newmathabbrev\BQPSPACE{\cfont{BQPSPACE}}
\newmathabbrev{\PCP}{\cfont{PCP}}
\newmathabbrev\BQUPSPACE{\cfont{BQ_UPSPACE}}
\newmathabbrev\QMAt{\QMA(2)}
\newmathabbrev\NQP{\cfont{NQP}}
\newmathabbrev\PreciseQMAt{\cfont{PreciseQMA}(2)}
\newmathabbrev\QMAtexp{\QMAt_{\exp}}
\newmathabbrev\MIP{\cfont{MIP}}
\newmathabbrev\MIPt{\MIP(2)}
\newmathabbrev\BellQMA{\cfont{BellQMA}}
\newmathabbrev\BellQMAt{\BellQMA(2)}
\newmathabbrev\BellQMAtexp{\BellQMAt_{\exp}}

\protected\def\verythinspace{%
  \ifmmode
    \mskip0.5\thinmuskip
  \else
    \ifhmode
      \kern0.08334em
    \fi
  \fi
}

\newcommand{\C}{\mathbb C}

\newcommand{\be}{\begin{equation}}
\newcommand{\ee}{\end{equation}}

\newcommand{\CNOT}{\mathrm{CNOT}}
\newcommand{\psiprod}{\psi_{\mathrm{prod}}}

\renewcommand{\epsilon}{\varepsilon}

\newcommand\lmin{\lambda_{\mathrm{min}}}
\newcommand\lmax{\lambda_{\mathrm{max}}}

\newcommand{\set}[1]{{\left\{#1\right\}}}    

\DeclareMathOperator{\poly}{poly}

\DeclareMathOperator{\Span}{Span}
\DeclareMathOperator{\trace}{Tr}

\DeclarePairedDelimiter\bra{\langle}{\rvert}
\DeclarePairedDelimiter\ket{\lvert}{\rangle}

\DeclarePairedDelimiter\abs{\lvert}{\rvert}
\DeclarePairedDelimiter\norm{\lVert}{\rVert}

\DeclarePairedDelimiterX\braket[2]{\langle}{\rangle}{#1 \delimsize\vert #2}
\DeclarePairedDelimiterX\ketbra[2]{\lvert}{\rvert}{#1 \delimsize\rangle\delimsize\langle #2}

\newcommand{\hin}{H_{\textup{in}}}
\newcommand{\hprop}{H_{\textup{prop}}}
\newcommand{\hstab}{H_{\textup{stab}}}

\newcommand{\spa}[1]{\mathcal{#1}}
\newcommand{\sA}{\spa{A}}

\newcommand{\pacc}{p_{\textup{accept}}}

\newcommand{\hout}{H_{\textup{out}}}

\newcommand{\base}{(\C^2)}
\newcommand{\Cn}{\base^{\otimes n}}
\newcommand{\Cm}[1]{\base^{\otimes #1}}

\newcommand{\ayes}{A_{\textup{yes}}} 
\newcommand{\ano}{A_{\textup{no}}} 
\newcommand{\ainv}{A_{\textup{inv}}} 
\newcommand{\psihist}{\psi_{\textup{hist}}}

\newcommand{\Hinit}{H_{\mathrm{init}}}
\newcommand{\pxy}{p_{x\rightarrow y}}
\newcommand{\psiinit}{\psi_{\mathrm{init}}}
\newcommand{\Hfinal}{H_{\mathrm{final}}}
\newcommand{\psifinal}{\psi_{\mathrm{final}}}

\newcommand{\lin}[1]{\textup{L}\left(#1\right)}

\newcommand{\unitary}[1]{\textup{U}\left(#1\right)}
\newcommand{\herm}[1]{\textup{Herm}\left(#1\right)}

\newcommand{\absA}{\abs{A}}

\newcommand{\absD}{\abs{D}}

\declaretheorem[numberwithin=section]{theorem}

\declaretheorem[sibling=theorem]{open question}

\declaretheorem[sibling=theorem,style=definition]{definition}

\declaretheorem[sibling=theorem,style=definition]{hint}

\title{The $7$ faces of quantum NP}
\date{}
\author{Sevag Gharibian\footnote{Department of Computer Science and Institute for Photonic Quantum Systems (PhoQS), Paderborn University, Germany. Email: sevag.gharibian@upb.de.}}

\begin{document}

\maketitle

\begin{abstract}
  When it comes to NP, its natural definition, its wide applicability across scientific disciplines, and its timeless relevance, the writing is on the wall: \emph{There can be only one.} Quantum NP, on the other hand, is clearly the apple that fell far from the tree of NP. 
  Two decades since the first definitions of quantum NP started rolling in, quantum complexity theorists face a stark reality: There's QMA, QCMA, $\QMAo$, $\QMAt$, StoqMA, and NQP. In this article aimed at a general theoretical computer science audience, I survey these various definitions of quantum NP, their strengths and weaknesses, and why most of them, for better or worse, actually appear to fit naturally into the complexity zoo.
\end{abstract}
\begin{quote}
    \emph{``Why, there's seven little chairs! Must be seven little children.''} --- Snow White
\end{quote}

\section{Introduction}\label{scn:intro}

The 1970's papers of Cook~\cite{cookComplexityTheoremprovingProcedures1971}, Levin~\cite{leonidlevinUniversalSearchProblems1973}, and Karp~\cite{karpReducibilityCombinatorialProblems1972} cemented NP as a staple of computer science curricula worldwide. Indeed, the average computer science graduate may not remember much about their {theoretical} computer science courses, but they most certainly remember that an NP-complete problem is one which \emph{should not be messed with}. This ability of NP to permeate a range of scientific disciplines stems from its exceedingly simple, yet natural definition. Looking for a class to capture what it means to ``efficiently verify a proof $x\in\set{0,1}^n$'' on a computer? For most practical purposes, NP is clearly it. Verify SAT solutions? Check. Authenticate your bank PIN at the ATM? Check. Confirm correctness of your Sudoku solution? \emph{Check.} In the words of Isaac Newton, ``Truth is ever to be found in simplicity, and not in the multiplicity and confusion of things'', a principle thoroughly exemplified by NP.

It is against this backdrop that we turn our attention to the ``ugly duckling'' of this story, the antithesis to the ``simplicity'' of NP, ``Quantum NP''. To set the stage, I encourage the reader to take a moment's pause, and imagine what a reasonable definition of ``quantum NP'' might be (this is where you stop reading, momentarily of course) --- odds are, you imagined the classical proof $x\in\set{0,1}^n$ replaced with a quantum proof $\ket{\psi}\in \Cn$, coupled with a now quantum verifier, presumably a poly-size quantum circuit, $V$. This is Quantum Merlin-Arthur (QMA), the \emph{de facto} definition of quantum NP. And it indeed captures the complexity of the canonical quantum generalization of Boolean Satisfiability, the \emph{Local Hamiltonian (LH)} problem (\Cref{scn:QMA}). However, whether this is the ``right'' definition is not entirely clear. For example, one can also feed the quantum verifier $V$ a \emph{classical} proof, $x\in\set{0,1}^n$ --- this yields Quantum-Classical Merlin-Arthur (QCMA) (\Cref{scn:QCMA}). Or, one can consider genuinely quantum phenomena, such as ``unentangled'' quantum proofs across a pair of spatially separated provers, yielding $\QMAt$ (\Cref{scn:QMAt}). In fact, there are so many variants of quantum NP, that one can name them after the seven dwarves of the classic tale, Snow White:
\begin{figure}[t]
  \begin{center}
    \begin{tabular}{c | c c c c c }
        Class/ & QMA/ & QCMA/  & $\QMAo$/ & $\QMAt$/ & StoqMA/ \\
       Nickname & Doc & Happy & Bashful & Grumpy & Dopey \\[0.5ex] 
       \hline      
       Complete  & LH & GSCON, & QSAT & Separable Sparse & Stoquastic LH \\
       problem & & VQA &&LH & \\
       \hline            
       Error & strong & strong & strong & weak & ? \\
       reduction &&&&&\\
       \hline            
       Universal & yes & yes & ? & yes & yes \\
       gate set &&&&&\\
       \hline            
       Perfect & ? & yes  &yes &? &? \\
       completeness &&&&&\\
       \hline            
       Hardness of & ? & yes &? &? &? \\
       approximation &&&&& \\
       \hline            
       Best known & $\AzPP\cap$ & $\QMAo$ &\QMA &$\QSigma_3\subseteq \NEXP$& $\SBP\subseteq \AM$ \\ 
       upper bound & $\pQMAlog$ & & & & \\
     \end{tabular}   
  \end{center}
  \caption{Variants of quantum NP. Note that NQP is omitted above, as it is not a true quantum analogue of NP, but rather a counting class (\Cref{scn:NQP}). Above, strong (weak) error reduction refers to amplification of a verification class' promise gap without (with) increasing the proof size.}
  \label{fig:classes}   
\end{figure}

\begin{figure}[t]
  \begin{center}
    \hspace{2.5cm}\includegraphics[width=9cm]{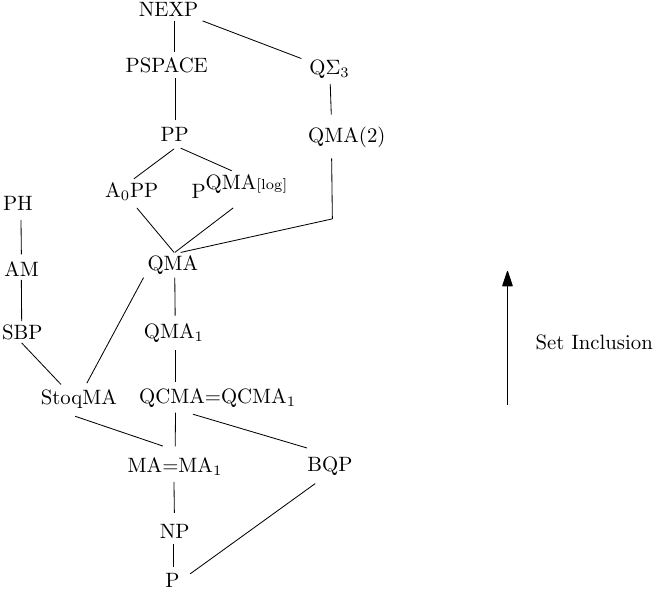}  
  \end{center}
  \caption{Class diagram for the various faces of quantum NP. Omitted again is NQP.}  
  \label{fig:classdiagram} 
\end{figure}
\begin{enumerate}
    \item (Doc) QMA: quantum proof and quantum verifier.
    \item (Happy) QCMA: classical proof and quantum verifier.
    \item (Bashful) $\QMAo$: QMA with perfect completeness.
    \item (Grumpy) $\QMAt$: QMA with two spatially separated provers.
    \item (Dopey) StoqMA: QMA with a verifier which can apply Hadamard gates in time step 1, followed by classical computation, then a single Hadamard on the output qubit immediately before measurement.
    \item (Sneezy) NQP: Quantum Turing machine which accepts with non-zero probability.
\end{enumerate}
As far as we know, all of these variants are distinct. Their properties are listed in \Cref{fig:classes}, and their relationship to one another in \Cref{fig:classdiagram}. The purpose of this article is to discuss each variant in a self-contained, accessible manner for a general theoretical computer science background. For clarity, this article is \emph{not} meant to be an exhaustive survey of all works on the topic --- that would not fit the size and scope of the present article. Rather, it aims to hopefully distill the ``main essence'' of each class, along with insights scattered throughout the literature. 

\paragraph{The $7$th dwarf.} The careful reader may have noticed we have only named $6$ dwarves above, whereas in Snow White, there are seven: This is our first open question.

\begin{open question}
    Establish the $7$th definition of quantum NP to take on the mantle of Sleepy.
\end{open question}
\begin{hint}
  Consider the possibility of \emph{quantum inputs}. Specifically, the definitions above take \emph{classical} inputs (i.e. a string describing the input). From a physical perspective, however, it is natural to consider inputs which are quantum states, $\ket{\psi}$ (see, e.g., the framework of classical shadows~\cite{aaronsonShadowTomographyQuantum2018,huangPredictingManyProperties2020}). A first step here is taken by Yamakami~\cite{yamakamiQuantumNPQuantum2002}.
\end{hint}
\begin{hint}
  Another option is the recent class $\cloneQMA$ of Nehoran and Zhandry~\cite{nehoranComputationalSeparationQuantum2023}, which is roughly QMA, but where the YES witness can be cloned\footnote{Formally, there exists a poly-time cloner $C$, such that for any YES input $x$, there exists a good witness $\ket{\psi}$ which additionally satisfies $C\ket{\psi}\approx \ket{\psi}\ket{\psi}$.}. Trivially, $\QCMA\subseteq\cloneQMA\subseteq\QMA$, and in fact there is a (unitary) oracle separation between $\QCMA$ and $\cloneQMA$~\cite{nehoranComputationalSeparationQuantum2023}.
\end{hint}

\paragraph{Organization.} We assume basic background in complexity theory, and try to assume as little quantum computing background as possible (a refresher is given in \Cref{scn:preliminaries}; for further detailed introductions to quantum computing and quantum complexity theory, course notes/videos are available~\cite{GharibianLabTeaching}). Sections are as follows: \Cref{scn:preliminaries} for preliminaries, \Cref{scn:QMA} for QMA, \Cref{scn:QCMA} for QCMA, \Cref{scn:QMAo} for $\QMAo$, \Cref{scn:QMAt} for $\QMAt$, \Cref{scn:StoqMA} for StoqMA, \Cref{scn:NQP} for NQP. 
\section{Preliminaries}\label{scn:preliminaries}

The conjugate transpose (i.e. adjoint) of matrix $A$ is $A^\dagger$, likewise for a column vector $\ket{\psi}$ is denoted $\bra{\psi}$. We use $\lin{\sA}$, $\herm{\sA}$, and $\unitary{\sA}$ to denote the sets of linear, Hermitian (i.e. $H=H^\dagger$), and unitary (i.e. $UU^\dagger=I$) operators acting on Hilbert space $\sA$, respectively.  The trace of matrix $A$ is $\trace(A)=\sum_i{A(i,i)}$, and the X-Men symbol $\otimes$ is the tensor product (for brevity, we sometimes omit this as in $X\otimes X=XX$ or $\ket{0}\otimes\ket{1}=\ket{01}$). We use $:=$ to denote a definition.

\paragraph{Quantum computing background.} A quantum state on $n$ qubits is specified by a unit vector $\ket{\psi}\in\Cn\simeq \C^{2^n}$. The set of allowed operations are the set of unitary operators $U\in \unitary{\Cn}$. Standard one-qubit gates used here are:
\begin{equation}\label{eqn:gates}
  X=\left(
    \begin{array}{cc}
      0 & 1 \\
      1 & 0 \\
    \end{array}
  \right)
  \quad 
  Y=\left(
    \begin{array}{cc}
      0 & -i \\
      i & 0 \\
    \end{array}
  \right)
\quad
  Z=\left(
    \begin{array}{cc}
      1 & 0 \\
      0 & -1 \\
    \end{array}
  \right)
  \quad
  H=\frac{1}{\sqrt{2}}\left(
    \begin{array}{cc}
      1 & 1 \\
      1 & -1 \\
    \end{array}
  \right)
  \quad
  T=\left(
    \begin{array}{cc}
      1 & 0 \\
      0 & e^{i\frac{\pi}{4}} \\
    \end{array}
  \right).
\end{equation}
The two-qubit Controlled-NOT gate
\begin{equation}
  \CNOT=\left(
    \begin{array}{cccc}
      1 & 0& 0 & 0 \\
      0 & 1 & 0 &0 \\
      0 & 0 & 0 & 1 \\
      0 & 0 & 1 & 0 \\
    \end{array}
  \right),
\end{equation}
together with Pauli $X$, $Y$, $Z$, Hadamard $H$, and $T$ are universal\footnote{The bottleneck for classical computers in simulating this gate set is the $T$ gate. Formally, simulating this gateset is Fixed-Parameter Tractable, i.e. poly-time in the number of Clifford gates (e.g. $X$, $Y$, $Z$, $H$, CNOT) but exponential in the number of $T$ gates, e.g.~\cite{aaronsonImprovedSimulationStabilizer2004,bravyiImprovedClassicalSimulation2016}.} for quantum computation. 

Measurements in this article are, without loss of generality, in the standard/computational basis $\set{\ket{0},\ldots, \ket{d-1}}\subset\C^d$. This is typically used in the study of quantum NP as follows: A verifier applies (1) a quantum circuit $V$ to (2) a proof $\ket{\psi}$ in register $A$ and initial state $\ket{0\cdots 0}$ in register $B$, and (3) measures a designated qubit of the result in the standard basis $\set{\ket{0},\ket{1}}$. The verifier accepts (rejects) if outcome $1$ ($0$) is obtained. Formally, defining single-qubit projector $\Pi=\ketbra{1}{1}$, the probability of outcome $1$ is given by\footnote{Note $\Pi$ acts non-trivially only a single qubit, whereas $\rho$ acts on $n$ qubits. Thus, to make dimensions formally match, the measurement operator is $\Pi\otimes I$ for $I$ acting on the $n-1$ qubits which $\Pi$ does not act on.\label{foot:local}} 
\begin{equation}
  \trace(\Pi \rho) \quad \text{ for }\quad \rho:=(V \ket{\psi}_A\ket{0\cdots 0}_B)(\bra{\psi}_A\bra{0\cdots 0}_BV^\dagger)\in\herm{\Cn}.
\end{equation}
Upon obtaining either outcome $\Pi$ (corresponding to $1$) or $I-\Pi$ (corresponding to $0$), our postmeasurement state collapses to $\Pi\rho\Pi/\trace(\Pi\rho)$ (respectively, $(I-\Pi)\rho(I-\Pi)/\trace((I-\Pi)\rho)$). 

\paragraph{Uniform circuit families and BQP.} Throughout, we use the notion of poly-time generated families of quantum circuits.

\begin{definition}[P-uniform quantum circuit family]\label{def:uniform}
  A family of quantum circuits $\set{V_n}$ is called P-uniform if there exists a polynomial-time Turing Machine $M$, which given as input $1^n$, outputs a classical description of $V_n$.
\end{definition}

The quantum generalization of P, or most accurately, of PromiseBPP, is defined as follows.

\begin{definition}[Bounded-Error Quantum Polynomial Time (BQP)]\label{def:BQP}
  A promise problem $\A=(\ayes,\ano)$ is in BQP if there exists a P-uniform quantum circuit family $\set{V_n}$ and polynomial $p:\N\rightarrow\N$ satisfying the following properties. For any input $x\in\set{0,1}^n$, $V_n$ takes in $n+p(n)$ qubits as input, consisting of the input $x$ on register $A$, and $p(n)$ ancilla qubits initialized to $\ket{0}$ on register $B$. The first qubit of register $B$, denoted $B_1$, is the designated output qubit, a measurement of which in the standard basis after applying $V_n$ yields the following:
  \begin{itemize}
      \item (Completeness) If $x\in \ayes$, $V_n$ accepts with probability $\geq 2/3$.
      \item (Soundness) If $x\in \ano$, then $V_n$ accepts with probability $\leq 1/3$.
  \end{itemize}
\end{definition}

\section{Doc: Quantum Merlin-Arthur (QMA)}\label{scn:QMA}

We begin with Doc, the leader of the dwarves, for which there is no better match than QMA. The \emph{de facto} definition of quantum NP, QMA was formalized by Kitaev (under the name BQNP)~\cite{kitaevClassicalQuantumComputation2002}:

\begin{definition}[Quantum Merlin Arthur (QMA)]\label{def:QMA}
  A promise problem $\A=(\ayes,\ano,\ainv)$ is in QMA if there exists a P-uniform quantum circuit family $\set{V_n}$ and polynomials $p,q:\N\rightarrow\N$ satisfying the following properties. For any input $x\in\set{0,1}^n$, $V_n$ takes in $n+p(n)+q(n)$ qubits as input, consisting of the input $x$ on register $A$, $p(n)$ qubits initialized to a quantum proof $\ket{\psi}\in\Cm{p(n)}$ on register $B$, and $q(n)$ ancilla qubits initialized to $\ket{0}$ on register $C$. The first qubit of register $C$, denoted $C_1$, is the designated output qubit, a measurement of which in the standard basis after applying $V_n$ yields the following:
  \begin{itemize}
      \item (Completeness) If $x\in \ayes$, $\exists$ proof $\ket{\psi}\in\Cm{p(n)}$ that $V_n$ accepts with probability $\geq 2/3$.
      \item (Soundness) If $x\in \ano$, then $\forall$ proofs $\ket{\psi}\in\Cm{p(n)}$, $V_n$ accepts with probability $\leq 1/3$.
      \item (Invalid case) If $x\in\ainv$, $V_n$ may accept or reject arbitrarily.
  \end{itemize}
\end{definition}
\noindent In words, QMA takes a poly-size quantum proof $\ket{\psi}$, and runs a poly-sized quantum verification circuit $V$. Note that QMA is a bounded-error \emph{promise} class\footnote{Actually, just about all quantum complexity classes are promise classes. Thus, to maintain sanity, we simply say (e.g.) ``QMA'' instead of ``PromiseQMA''.}, not a language! Thus, it most accurately generalizes PromiseMA (i.e. the bounded-error promise version of NP). While this makes the class ``nice'', in that it has complete problems, it can make handling \emph{oracle} calls to QMA for classes such as $\pQMAlog$ a serious obstacle~\cite{ambainisPhysicalProblemsThat2014a,gharibianComplexitySimulatingLocal2018} --- this is why we explicitly stress the existence of the ``invalid case'' in \Cref{def:QMA}. Finally, the completeness and soundness parameters above can be improved to $1-1/2^n$ and $2^n$, respectively, via two methods: \emph{Weak} error reduction, which applies standard parallel repetition and thus blows up the proof size~\cite{aharonovQuantumNPSurvey2002}, and \emph{strong} error reduction, which remarkably does \emph{not} need to increase the proof size~\cite{marriottQuantumArthurMerlin2005}.

\paragraph{\emph{The} complete problem.} While QMA has a reasonable number of complete problems~\cite{bookatzQMAcompleteProblems2014}, its raison d'etre is arguably that it captures the physically motivated quantum generalization of Boolean Satisfiability, the \emph{Local Hamiltonian (LH)} problem. \\

\noindent\emph{Intuition.} To motivate LH, consider the NP-complete problem MAX CUT, in which given a simple undirected graph $G=(V,E)$ on $n$ vertices, one wishes to assign a label from $\set{0,1}$ to each vertex, so that as many edges as possible have distinct labels. Formally, this can be modelled via a minimum-eigenvalue problem as follows. The matrix 
\begin{equation}\label{eqn:ZZ}
  Z\otimes Z=\left(
    \begin{array}{cccc}
      1 & 0& 0 & 0 \\
      0 & -1 & 0 &0 \\
      0 & 0 & -1 & 0 \\
      0 & 0 & 0 & 1 \\
    \end{array}
  \right)
\end{equation}
has eigenvalues $\set{1,-1,-1,1}$, with corresponding eigenvectors $\set{\ket{00}, \ket{01}, \ket{10}, \ket{11}}$. Thus, the eigenspace corresponding to the minimal eigenvalue, called the \emph{ground space}, is $\Span(\ket{01},\ket{10})$. But this is precisely the labelling we wish to see on each satisfied edge of MAX CUT --- either $01$ or $10$. By placing this constraint on each edge of the ``interaction graph'' $G$, we obtain the local Hamiltonian $H=\sum_{(i,j)\in E}Z_i\otimes Z_j$ encoding the same MAX CUT instance\footnote{Two subtleties: (1) Formally, this holds since all constraints $Z_i\otimes Z_j$ are diagonal in the same basis, the standard basis. Thus, without loss of generality, the ground state (i.e. minimal eigenvector) is also a standard basis state $\ket{x}\in\Cn$ for some string $x\in\set{0,1}^n$, which may be viewed as encoding a cut in $G$. (2) Technically, this encodes a shifted version of MAX CUT, in which each satisfied edge has value $-1$, and each unsatisfied edge has value $1$. This is easily converted to $1$ and $0$, respectively, by considering $(I+ZZ)/2$ and seeking the \emph{maximum} eigenvalue.}. Thus, MAX CUT is reduced to a minimum eigenvalue problem, i.e. estimating $\lmin(H)$. The catch? $H$ acts on $n$ qubits, and thus has dimension $2^n\times 2^n$, meaning brute force diagonalization to extract $\lmin(H)$ takes exponential time. In sum, we have encoded an NP-hard problem into estimating $\lmin(H)$.

Next, can we encode anything harder into $\lmin(H)$? Observe that our constraints (\Cref{eqn:ZZ}) are diagonal in the \emph{standard} basis --- what happens if we deviate from this basis? For example, a standard operator basis for $\herm{\C^2}$ is $\set{I,X,Y,Z}$ i.e. any $M\in \herm{\C^2}$ can be written $M=aI+bX+cY+dZ$ for $a,b,c,d\in\R$. Thus far, we have only used $I$ and $Z$ in constructing $H$ --- what happens if we throw in $X$ and $Y$, which are \emph{not} diagonal in the standard basis? The resulting Hamiltonian, 
\begin{equation}\label{eqn:Heisenberg}  
  H=\sum_{(i,j)\in E} X_i\otimes X_j+Y_i\otimes Y_j+ Z_i\otimes Z_j = \sum_{(i,j)\in E} (XX+YY+ZZ)_{ij},
\end{equation}
is the Heisenberg anti-ferromagnet, a notoriously difficult to solve\footnote{When a physicist speaks of ``solving'' a model, they generally mean solving for some context-dependent property of interest. Here, we mean solving for the ground state energy.} model from the study of quantum magnetism dating back over a century. Formally, each local term $XX+YY+ZZ$ has a unique two-qubit ground state, the maximally entangled \emph{singlet} state $(\ket{01}-\ket{10})/\sqrt{2}$. This looks \emph{a lot} like a satisfying MAX CUT assignment on an edge, earning this model the monikker of ``Quantum MAX CUT''\footnote{For clarity, by Quantum MAX CUT, one refers to having anti-ferromagnetic constraints of form $\alpha_{ij}(XX+YY+ZZ)_{ij}$ for $\alpha_{ij}\geq 0$. When $\alpha_{ij}=1$ for all $(i,j)\in E$, this model is called \emph{the} Heisenberg anti-ferromagnet. We remark SDP-based approximation algorithm techniques as in the Goemans-Williamson MAX CUT algorithm~\cite{goemansImprovedApproximationAlgorithms1995} also apply to Quantum MAX CUT~\cite{gharibianAlmostOptimalClassical2019,parekhOptimalProductStateApproximation2022}.}~\cite{gharibianAlmostOptimalClassical2019}. Just as MAX-CUT is NP-complete, Quantum MAX CUT is QMA-complete~\cite{cubittComplexityClassificationLocal2016a,piddockComplexityAntiferromagneticInteractions2017,cubittUniversalQuantumHamiltonians2018a}, albeit with poly-size weights on the edges.\\ 

\noindent \emph{The general $k$-local Hamiltonian ($k$-LH) problem.} With the intuition in place, we can state the most general version of $k$-LH. Here, the input is an $n$-qubit Hermitian matrix $H$ with succinct description $H=\sum_{S\subset[n]} H_S $, where each local ``quantum clause'' $H_S$ acts on some subset $S$ of $k$ qubits. Given threshold parameters $\alpha,\beta\in\R$ satisfying $\beta-\alpha\geq 1/\poly(n)$, the goal is to output YES if $\lmin(H)\leq \alpha$ or NO if $\lmin(H)\geq \beta$. Physically, $\lmin(H)$ is the energy level (i.e. ground state energy) to which the quantum many-body system described by $H$ relaxes when cooled to near absolute zero (think back to high school chemistry --- remember how electrons like to settle into their lowest energy configuration?). The problem $k$-LH is thus strongly physically motivated, being a central focus in quantum chemistry~\cite{leeThereEvidenceExponential2022}.

Of course, that's all fine and good for the physicists, but why should we as computer scientists care? Simple: $k$-LH is to QMA as\footnote{The MAX in MAX-$k$-SAT is important here --- $k$-SAT is instead more closely related to $\QMAo$ (\Cref{scn:QMAo}).} MAX-$k$-SAT is to NP --- it is \emph{the} canonical QMA-complete problem. Kitaev was the first~\cite{kitaevClassicalQuantumComputation2002} to show that $5$-LH is QMA-complete. This was improved to QMA-hardness for $2$-LH~\cite{kempejulia3localHamiltonianQMAcomplete2003,kempeComplexityLocalHamiltonian2006} via the introduction of \emph{perturbation theory gadgets}, which opened Pandora's box. With the latter in hand, QMA-completeness on the 2D lattice followed~\cite{oliveirarobertoComplexityQuantumSpin2008}, along with a quantum analogue~\cite{cubittComplexityClassificationLocal2016a,bravyiComplexityQuantumIsing2017} of Shaefer's dichotomy theorem~\cite{schaeferComplexitySatisfiabilityProblems1978}: Whereas SAT problems are either in P or NP-complete, $k$-LH problems satisfy a ``quad''-chotomy theorem --- they are either in P, NP-complete, StoqMA-complete (Dopey, \Cref{scn:StoqMA}), or QMA-complete, depending on the family of local constraints permitted. Included in the QMA-complete portion of this classification are the Heisenberg anti-ferromagnetic ($XX+YY+ZZ$, \Cref{eqn:Heisenberg}) and XY (i.e. $XX+YY$) interactions. Finally, QMA-hardness holds \emph{even on a 1D chain}, i.e. where all constraints are $2$-local and act on neighboring sets $\set{i,i+1}$ on the line~\cite{aharonovPowerQuantumSystems2009,nagajLocalHamiltoniansQuantum2008,hallgrenLocalHamiltonianProblem2013} (albeit with local dimension $d=8$). This is in strong contrast to MAX-$k$-SAT on the line, which can be efficiently solved via dynamic programming. In fact, even if all 1D constraints are \emph{identical}, i.e. the \emph{translationally invariant} setting, the problem remains hard\footnote{Formally, one obtains hardness for QMAEXP, which is to QMA as NEXP is to NP. This is because in the translation invariant setting, the only input to the problem is the length of the chain, $n$, specified in binary. The interaction term $H_{i,i+1}$ repeated along the chain encodes the quantum Turing machine verifying the QMAEXP problem. Since this Turing machine is independent of the input, the local dimension of $H_{i,i+1}$ is constant.}~\cite{gottesmanQuantumClassicalComplexity2009,bauschComplexityTranslationallyInvariant2017} (at the cost of increasing the local dimension to a larger constant). \\

\noindent \emph{The fine print.} Let us take a moment to discuss some finer points regarding the results above.
\begin{itemize}
    \item \emph{Perturbation theory gadgets.} While powerful, these gadgets have a distinct drawback --- they require the placement of poly-sized real weights on the local terms of the Hamiltonian. This is physically not so well-motivated --- when a physicist refers to ``the Heisenberg anti-ferromagnet'', for example, they typically mean with \emph{unit} weights. In this case, we do \emph{not} have a proof of hardness for the anti-ferromagnet --- not even NP-hardness!
    \begin{open question}
      Prove that that Quantum MAX CUT with unit edge weights is QMA-hard. Failing that, prove NP-hardness.
    \end{open question}
    \item \emph{Is Quantum MAX CUT really the quantum generalization of MAX CUT?} As the Germans would say, \emph{jein} (meaning ``yes and no'', or ``ja und nein''). On the one hand, as per \Cref{eqn:Heisenberg}, both MAX CUT (MC) and Quantum MAX CUT (QMC) clearly belong to the same family of Hamiltonians. On the other hand, we do not know of a \emph{direct} embedding of MC into QMC, other than to go via the indirect route of perturbation theory gadgets~\cite{cubittComplexityClassificationLocal2016a}. Moreover, while MC on bipartite graphs is poly-time solvable (i.e. it is a $2$-coloring problem), there are very few known bipartite graphs on which QMC can be efficiently solved, e.g. the 1D chain (via the Bethe ansatz~\cite{betheZurTheorieMetalle1931}), and the complete bipartite graph (e.g.~\cite{cubittComplexityClassificationLocal2016a}) (more recently, see~\cite{wattsRelaxationsExactSolutions2023a,takahashiSUSymmetricSemidefinite2023}). Nevertheless, QMC on bipartite graphs is \emph{not} expected to be QMA-hard, as in this setting, it falls~\cite{cubittComplexityClassificationLocal2016a} into $\StoqMA\subseteq\QMA$ (\Cref{scn:StoqMA}).
    \begin{open question}
      Is there a ``clean/direct'' NP-hardness reduction from MC to QMC?
    \end{open question}
    \begin{open question}
      What is the complexity of QMC on bipartite graphs?
    \end{open question}
  \end{itemize}

\paragraph{Circuit-to-Hamiltonian constructions.} A central tool in proving QMA-hardness results for LH is Kitaev's~\cite{kitaevClassicalQuantumComputation2002} circuit-to-Hamiltonian construction, which ``quantizes'' the Cook-Levin construction~\cite{cookComplexityTheoremprovingProcedures1971,leonidlevinUniversalSearchProblems1973}. For this, we consider any quantum verifier $V=V_L\cdots V_1$ consisting of two-qubit gates $V_i$, applied to initial state $\ket{\phi}_A\ket{0\cdots 0}_B$, where $\ket{\phi}$ is a quantum proof and $B$ is the ancilla. Our goal is to track the sequence of ``quantum configurations'' $\ket{\psi_t}$ encountered by $V$ over time, and use ``local Hamiltonian checks'' to ensure the propagation from configuration $\ket{\psi_t}$ to $\ket{\psi_{t+1}}$ proceeds correctly. However, in contrast to Cook-Levin, which encodes each configuration as a row of a tableau, we encode configurations in \emph{superposition}, i.e. as $\sum_t\ket{\psi_t}$. In doing so, however, we lose our notion of \emph{time}, in that for a tableau, time was encoded by \emph{position/row index}. To recover this, Kitaev used an idea of Feynman~\cite{feynmanQuantumMechanicalComputers1986} and attached a new ancilla register to track time, $C$, denoted the ``clock'' register. The resulting ``quantum tableau'' is known as a \emph{history state},
\begin{equation}\label{eqn:hist}
    \ket{\psihist}=\frac{1}{\sqrt{L+1}}\sum_{t=0}^L V_t\cdots V_1\ket{\psi}_A\ket{0\cdots 0}_B\ket{t}_C.
\end{equation}
The local checks enforcing this history state structure are now as follows:
\begin{itemize}
  \item (Ancilla initialization: $\hin$) Enforces that at time $t=0$, ancilla register $B$ is all zeroes:
  \begin{equation}
      \hin = I_A\otimes \left(\sum_i \ketbra{1}{1}_{B_i}\right)\otimes \ketbra{0}{0}_C.
  \end{equation}
  Intuitively, when the clock reads $0$, an energy \emph{penalty} is administered if any bit of $B$ has overlap with $\ket{1}$.

  \item (Correct output: $\hout$.) Checks whether, at time $L$, the verifier accepts (we assume $B_1$ is the output qubit):
  \begin{equation}
    \hout = I_A\otimes \ketbra{0}{0}_{B_1}\otimes\ketbra{L}{L}_C.
  \end{equation}
  Intuitively, when the clock reads $L$, an energy penalty is administered if the verifier's output qubit has overlap with $\ket{0}$, i.e. with the ``REJECT'' outcome.

  \item (Correct propagation: $\hprop$.) Ensures configuration $t+1$ follows from configuration $t$:
  \begin{equation}\label{eqn:hprop}
  \hprop = \sum_{t=0}^{L-1} -V_{t+1}\otimes\ketbra{t+1}{t}_C - V_{t+1}^\dagger\otimes\ketbra{t}{t+1}_C,
  \end{equation}
  where recall $V_t$ acts on $A$ (proof) and $B$ (ancilla). The first term ensures that, in going from time $t$ to $t+1$, we apply $V_{t+1}$. (The second term ensures $\hprop$ is Hermitian.)
\end{itemize}
Technically, there is also a fourth set of checks, $\hstab$, which ensures the clock $C$ is correctly encoded in unary (this allows all terms in our Hamiltonian to be $5$-local, as opposed to $\log$-local if we had used a binary clock encoding); we omit this for brevity. In sum, the output of the construction is $5$-local Hamiltonian $H=\hin+\hprop+\hout+\hstab$, which can be shown to satisfy: If there exists a proof $\ket{\psi}$ accepted by $V$ with probability at least $1-\epsilon$, then the history state obtains $\bra{\psihist}H\ket{\psihist}\in O(\epsilon/L)$. If, conversely, the best proof is accepted with probability at most $\epsilon$, then $\lmin(H)\in\Omega((1-\sqrt{\epsilon})/L^3)$. Since recall the completeness and soundness error $\epsilon$ can be exponentially reduced for QMA, we obtain a promise gap scaling as $1/L^3$. This has since been improved to $1/L^2$~\cite{bauschAnalysisLimitationsModified2018,cahaClocksFeynmanComputer2018a,watsonDetailedAnalysisCircuittoHamiltonian2019}.

\begin{open question}
    The strength of Kitaev's construction is its generality --- it is agnostic to the value of $L$ (which could even be exponential, at the cost of having $\hprop$ have exponentially many local terms, e.g.~\cite{gharibianQuantumSpaceGround2023}), the type of proof in $A$, or even if there is a proof at all (e.g.~\cite{wocjanSeveralNaturalBQPComplete2006,harrowQuantumAlgorithmLinear2009,gharibianDequantizingQuantumSingular2022}). A weakness, however, is that one requires error reduction for $V$ (i.e. $\epsilon\in o(L)$) in order to obtain a non-empty promise gap, which for example is not known to hold for StoqMA~\cite{aharonovStoqMAVsMA2021} (\Cref{scn:StoqMA}). Is there a circuit-to-Hamiltonian construction which works for $\epsilon\in\Theta(1)$?
\end{open question}

\paragraph{Promise gaps and hardness of approximation.} Kitaev's construction shows LH is QMA-hard for inverse polynomial promise gap $\beta-\alpha\geq 1/\poly(n)$. What about different promise gaps?\\

\noindent\emph{Inverse exponential gap.} In the case of inverse \emph{exponential} promise gap, the corresponding ``Precise Local Hamiltonian (PreciseLH)'' problem is complete for PreciseQMA, defined as QMA with exponentially small promise gap. The latter, surprisingly, equals PSPACE~\cite{feffermanCompleteCharacterizationUnitary2018a}. \emph{Why} is this surprising? Simply because NP and QMA behave drastically different in this setting. Consider MAX-SAT with exponentially large clause weights, which we shall call PreciseSAT. Deciding whether an instance of PreciseSAT has optimal value $2^n$ versus $2^n-1$ is in NP, since $n$-bit arithmetic can be done in $\poly(n)$-time on a Turing machine. Via rescaling, PreciseLH can \emph{also} be written this way, i.e. as LH with a \emph{constant} promise gap and \emph{exponential} weights on quantum clauses. Yet, the latter is PSPACE-hard! Morally, this is because a QMA machine is a \emph{sampling} device, with the optimal ``value'' of its objective function (say, energy penalty against a given state $\ket{\psi}$) encoded in its \emph{probability} $\pacc$ of acceptance. And the Chernoff bound tells us that, via polynomially many independent runs of a sampling experiment, we can only approximate $\pacc$ within additive inverse \emph{polynomial} error. \\

\noindent\emph{Constant gap/hardness of approximation.} The case of \emph{constant} promise gap for LH is the regime of the fabled Quantum PCP conjecture~\cite{aharonovGuestColumnQuantum2013,aharonovQuantumPCPConjecture2013}. To make ``constant'' promise gap well-defined, one typically rescales the entire Hamiltonian $H$ so that its spectral norm satisfies $\norm{H}\leq 1$. We then ask: Is there a constant $c$, such that given a (positive semi-definite) local Hamiltonian $H$ with $\norm{H}\leq 1$, it is QMA-hard to decide whether $\lmin(H)=0$ or $\lmin(H)\geq c$? 

While the conjecture remains very much open, there has been some exciting recent progress. If one believes $\QCMA\neq \QMA$ (i.e. that classical proofs are weaker than quantum proofs for a quantum verifier), then as a necessary requirement for Quantum PCP, there must exist a family of local Hamiltonians $H$ whose low energy space (up to some constant, $c$) is spanned by states requiring \emph{superpolynomial}-size quantum circuits to prepare. For if a sub-$c$ energy state \emph{could} be prepared by poly-size quantum circuit $V$, and \emph{if} the quantum PCP conjecture holds for $c$, then a QCMA prover could send a classical description of $V$ to a QMA prover, collapsing $\QCMA=\QMA$. As an aside, this says nothing about whether $H$ encodes a QMA computation --- it simply says the ``$c$-low'' energy space  of $H$ is ``complicated''. 

Now, as a field, we very much regret that even showing that this basic requirement holds remains open. However, its younger sibling, the No Low-Energy Trivial States (NLTS) conjecture~\cite{freedmanQuantumSystemsNonkhyperfinite2014}, \emph{has} recently been shown~\cite{anshuNLTSHamiltoniansGood2023}!  In NLTS, one replaces ``superpolynomial-size'' quantum circuit with ``superconstant-depth'' quantum circuit. Thus, if Quantum PCP holds and NLTS does not, then we get an even stronger collapse, $\NP=\QMA$. This is because a standard light cone argument shows that, given $k$-local Hamiltonian $H$ with $k\in O(1)$ and constant depth quantum circuit $U$, one can efficiently \emph{classically} evaluate $\bra{0\cdots 0}U^\dagger H U\ket{0\cdots 0}$. Now that NLTS is shown, we ask, \emph{what is the next step towards quantum PCP}? One option is to take inspiration from the recent study of \emph{dequantization} in quantum algorithms~\cite{tangQuantuminspiredClassicalAlgorithm2019,chiaSamplingbasedSublinearLowrank2020} and define a stronger conjecture, such as the No Low-Energy Samplable States (NLSS) conjecture~\cite{gharibianDequantizingQuantumSingular2022} (see also the NLCES conjecture~\cite{weggemansGuidableLocalHamiltonian2023}). This replaces ``states preparable with a constant depth quantum circuit'' in NLTS with ``states allowing classical sampling access''\footnote{Roughly, we have sampling access~\cite{tangQuantuminspiredClassicalAlgorithm2019} to $\ket{\psi}=\sum_i\alpha_i\ket{i}\in\Cn$ if we can efficiently (1) compute any $\alpha_i$ given index $i$, (2) sample index $i$ with probability $\abs{\alpha_i}^2$ (i.e. simulate measurement of $\ket{\psi}$ in the standard basis), and (3) estimate the norm of $\ket{\psi}$.}. A special case of states with sampling access for which NLSS was recently confirmed~\cite{cobleLocalHamiltoniansNo2023} is stabilizer states, i.e. those preparable only with Clifford gates. In fact, we now know of Hamiltonians whose low energy space requires $\Omega(\log n)$ depth and at least $\Omega(n)$ $T$-gates (i.e. non-stabilizer gates) to prepare~\cite{cobleHamiltoniansWhoseLowenergy2023}.

\begin{open question}
    The NLTS constructions above leverage recent breakthroughs~\cite{dinurLocallyTestableCodes2022,panteleevAsymptoticallyGoodQuantum2022, leverrierQuantumTannerCodes2022} in the study of quantum LDPC codes. Extending these NLTS results to a full quantum PCP, however, faces a major barrier --- the codes used are stabilizer codes~\cite{anshuNLTSHamiltoniansGood2023,leverrierQuantumTannerCodes2022}, which cannot encode QMA-complete problems unless $\p=\QMA$~\cite{yanKlocalPauliCommuting2012}. As a first step, can computations weaker than QMA be embedded in variations of such constructions, such as NP-hard computations?
\end{open question}

\paragraph{Upper bounds.} The best known upper bound on QMA is $\AzPP\cap\pQMAlog$. The first class, $\AzPP\subseteq\PP$~\cite{vyalyiQMAPPImplies2003}, is defined as PP, except where the completeness/soundness parameters $\alpha$ and $\beta$ satisfy $\alpha\geq 2\beta$ (note both $\alpha$ and $\beta$ can be exponentially small). Note that we do not believe $\AzPP=\PP$, as otherwise $\PH\subseteq\PP$~\cite{vyalyiQMAPPImplies2003} for $\PH$ the Polynomial-Time Hierarchy~\cite{stockmeyerPolynomialtimeHierarchy1976}. The second class, $\pQMAlog\subseteq \PP$~\cite{ambainisPhysicalProblemsThat2014a, gharibianComplexitySimulatingLocal2019a}, is the set of decision problems solved by a P machine  making at most $O(\log n)$ queries to a QMA oracle. It is unlikely that $\QMA=\pQMAlog$, as $\coQMA\subseteq\pQMAlog$.
\begin{open question}
  $\AzPP$ and $\pQMAlog$ are \emph{complementary}, in the sense that $\AzPP$ can handle exponentially small promise gaps, but not universal quantifiers (as in $\coQMA$), whereas $\pQMAlog$ handles universal quantifiers but not small promise gaps. Intuitively, their intersection should thus capture \emph{neither} of these two extreme properties. So, is $\QMA=\AzPP\cap\pQMAlog$? 
\end{open question}

\section{Happy: Quantum-Classical Merlin-Arthur (QCMA)}\label{scn:QCMA}

In a world where everyone wants the latest and the greatest, Quantum-Classical Merlin-Arthur is chill --- no need for a fancy quantum proof, a classical proof will do just fine. Life is short, be \emph{happy}, as Mamma QCMA used to say. First formalized by Aharonov and Naveh~\cite{aharonovQuantumNPSurvey2002}, QCMA sits naturally between MA and QMA, and is defined as follows.

\begin{definition}[Quantum-Classical Merlin Arthur (QCMA)]\label{def:QCMA}
  A promise problem $\A=(\ayes,\ano)$ is in QCMA if there exists a P-uniform quantum circuit family $\set{V_n}$ and polynomials $p,q:\N\rightarrow\N$ satisfying the following properties. For any input $x\in\set{0,1}^n$, $V_n$ takes in $n+p(n)+q(n)$ qubits as input, consisting of the input $x$ on register $A$, $p(n)$ qubits initialized to a classical proof $\ket{y}\in\set{0,1}^{p(n)}$ on register $B$, and $q(n)$ ancilla qubits initialized to $\ket{0}$ on register $C$. The first qubit of register $C$, denoted $C_1$, is the designated output qubit, a measurement of which in the standard basis after applying $V_n$ yields the following:
  \begin{itemize}
      \item (Completeness) If $x\in \ayes$, $\exists$ proof $y\in\set{0,1}^{p(n)}$ that $V_n$ accepts with probability $\geq 2/3$.
      \item (Soundness) If $x\in \ano$, then $\forall$ proofs $y\in\set{0,1}^{p(n)}$, $V_n$ accepts with probability $\leq 1/3$.
  \end{itemize}
\end{definition}
\noindent The fact that QCMA contents itself with a \emph{classical} proof means two things: (1) QCMA-hardness doesn't seem to capture any ``nice'' family of Hamiltonians (in the sense of how, for example, the anti-ferromagnetic Heisenberg interaction was QMA-complete). This is intuitively because when a QCMA verifier $V$ is pushed through Kitaev's circuit-to-Hamiltonian mapping~\cite{wocjanTwoQCMAcompleteProblems2003}, the resulting Hamiltonian has a\footnote{This is because the history state (\Cref{eqn:hist}) when $\ket{\psi}$ is a classical string $y$ has a poly-size preparation circuit. Of course, \emph{finding} this circuit is QCMA-hard, since it requires finding $y$ to begin with.} low-energy state preparable by a poly-size circuit. And it is not clear whether this latter property should manifest itself in any ``syntactic'' sense in terms of the family of local terms the Hamiltonian is allowed to use. Indeed, the careful reader may have noticed that QCMA is consipiciously absent from the quantum ``quad''-chotomy theorem for LH~\cite{cubittComplexityClassificationLocal2016a,bravyiComplexityQuantumIsing2017}, which is based on precisely such a ``syntactic'' characterization of local terms. (2) In return, QCMA gives us other goodies not known to hold for QMA --- perfect completeness and hardness of approximation.

\paragraph{Perfect completeness.} That QCMA admits strong error reduction, i.e. without blowing up the proof size, is trivial since the proof is now classical, and thus can be copied for multiple parallel runs of the verifier. What is less obvious is that, like MA and \emph{unlike} BQP (the quantum generalization of BPP), QCMA admits perfect completeness~\cite{jordanAchievingPerfectCompleteness2012}. The kernel of this result is the observation that if one can make $V$ accept in the YES case with a ``nice'' probability such as $1/2$, then amplitude amplification~\cite{chiQuantumDatabaseSearch1999,brassardQuantumAmplitudeAmplification2002} or quantum rewinding~\cite{watrousZeroKnowledgeQuantumAttacks2009} can be used to boost this acceptance probability precisely to $1$. To achieve this $1/2$, two ingredients are used: (1) By choosing an appropriate universal gate set for QCMA consisting only of entries in $\set{0,1,1/\sqrt{2}}$, namely using the Hadamard, Toffoli, and $X$ gate set~\cite{shiBothToffoliControlledNOT2003,aharonovSimpleProofThat2003}, one may assume that any acceptance probability of $V$ on a classical proof $y$ is of form $k/2^{\poly(n)}$ for integer $k$. Thus, the honest prover can send \emph{both} $y$ and $k$. (2) Given $k$, the verifier now knows (in the honest case) how to ``rebalance'' its verification so that on proof $y$, it accepts with probability $1/2$. This is roughly done by running, in equal superposition,  the original verification $V$ on $y$, and a random sampling of an integer $z$ between $1$ and $2^{\poly(n)}$ and accepting if $z\geq k$. Finally, note this strategy fails for BQP and QMA, since for the former, $k$ cannot be sent as a proof, and for the latter, the quantum proof $\ket{\psi}$ can break the fact that acceptance probabilities are of form $k/2^{\poly(n)}$.

\paragraph{Complete problems.} Moving to complete problems, the itch that QCMA needs to scratch is --- \emph{What good is a \emph{classical} proof to a \emph{quantum} verifier?} This is one of those hard-to-reach itches, and so pickings for QCMA-complete problems are slimmer than for QMA (e.g.~\cite{wocjanTwoQCMAcompleteProblems2003,wocjanJonesPolynomialQuantum2008,janzingBQPcompleteProblemsConcerning2006,weggemansGuidableLocalHamiltonian2023}). Here, we focus on two natural problems: Ground State Connectivity (GSCON), which we discuss now, and Minimization for Variational Quantum Algorithms (MIN-VQA), which we subsequently discuss under ``Hardness of approximation''.

The Ground State Connectivity problem  (GSCON)~\cite{gharibianGroundStateConnectivity2015,gossetQCMAHardnessGround2017} is closely related to LH, except instead of asking for a Hamiltonian $H$'s ground state energy, it asks whether the low-energy landscape of $H$ has an \emph{energy barrier}. More formally, as input one is given two ground states $\ket{\psi_1}$ and $\ket{\psi_2}$ (specified via quantum circuits) of a local Hamiltonian $H$, and the goal is to decide: Does there exist a poly-length sequence of $2$-qubit gates $(U_1,U_2,\cdots , U_m)$ satisfying (1) $U_m\cdots U_1\ket{\psi_1}\approx\ket{\psi_2}$, and (2) for all $i\in[m]$, $U_i\cdots U_1\ket{\psi_1}$ is low energy against $H$? Note that containment of GSCON in QCMA is clear --- the prover sends the classical description of local gates $(U_1,\ldots, U_m)$, and the verifier checks condition (1) using the SWAP test~\cite{buhrmanQuantumFingerprinting2001} (\Cref{fig:ProductStatetest} in \Cref{scn:QMAt}) and condition (2) via (e.g.) quantum phase estimation~\cite{kitaevQuantumMeasurementsAbelian1995a}. 

QCMA-hardness of GSCON turns out to be quite robust. First, QCMA-hard holds even if the local terms of $H$ pairwise commute~\cite{gossetQCMAHardnessGround2017}. This is in contrast to LH with commuting terms, for which no QMA-hardness proof is known, and moreover which is in NP for (e.g.) $2$-local commuting Hamiltonians~\cite{bravyiCommutativeVersionLocal2005,aharonovComplexityCommutingLocal2011,schuchComplexityCommutingHamiltonians2011,aharonovComplexityTwoDimensional2018,iraniCommutingLocalHamiltonian2023}! 
\begin{open question}
  What is the complexity of commuting $k$-LH for general $k\in O(1)$?
\end{open question}
\noindent Second, GSCON remains hard (in this case, QCMAEXP-hard) in the $1$D translation invariant setting on a chain of length $n$, \emph{even if} instead of $2$-qubit gates, one allows $(n-1)$-local\footnote{Note that allowing $n$-local gates trivializes the problem, as clearly there exists an $n$-local unitary mapping $\ket{\psi_1}$ to $\ket{\psi_2}$. Thus, soundness against $(n-1)$-local gates is optimal.} gates~\cite{watsonComplexityTranslationallyInvariant2023b}.\\ 
  
\noindent\emph{Traversal Lemma.} All known proofs of QCMA-hardness for GSCON leverage a simple underlying principle --- take the QCMA verifier $V$, map it to a local Hamiltonian $H'$ via circuit-to-Hamiltonian mapping, and embed $H'$ into a triple $(H,\ket{\psi_1},\ket{\psi_2})$, so that going from $\ket{\psi_1}$ to $\ket{\psi_2}$ through $H$'s low energy space forces one to ``switch on'' $H'$. In its simplest form, one sets~\cite{gharibianGroundStateConnectivity2015} 
\begin{align}
    H&=H'_A\otimes (I-\ketbra{000}{000}-\ketbra{111}{111})_S\text{ for $S$ the ``switch'' register},\label{eqn:switch}\\
    \ket{\psi_1}&=\ket{0\cdots 0}_A\ket{000}_S,\label{eqn:psi1}\\
    \ket{\psi_2}&=\ket{0\cdots 0}_A\ket{111}_S.\label{eqn:psi2}
\end{align}
The intuition is as follows: Starting with $\ket{\psi_1}$, the honest prover (1) prepares the history state for $H'$ in $A$, then (2) flips $\ket{000}_S$ to $\ket{111}$ via Pauli $X$ gates, and finally (3) uncomputes the history state in $A$ to obtain $\ket{\psi_2}$. Since all gates applied must be $2$-local, in step 2 the prover's state has non-zero overlap with $(I-\ketbra{000}{000}-\ketbra{111}{111})_S$ in \Cref{eqn:switch}, which ``switches on'' $H'$, which checks the history state in $A$. 

\begin{figure}[t]
  \begin{center}
    \includegraphics[width=4cm]{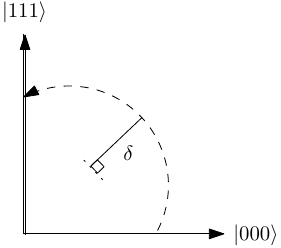}
  \end{center}
  \caption{A special case of the Traversal Lemma. Here, $\ket{000}$ and $\ket{111}$, which are clearly $2$-orthogonal. The lemma says any sequence of $m$ $2$-local gates (depicted by the dotted arc) mapping $\ket{000}$ $\epsilon$-close to $\ket{111}$ must leave the 2D span of $\ket{000}$ and $\ket{111}$ by $\delta\geq (1-\epsilon)/m$. } 
  \label{fig:traversallemma}
\end{figure}

The crucial question now is soundness --- how do we know there isn't some \emph{other} sequence of $2$-local gates mapping $\ket{\psi_1}$ to $\ket{\psi_2}$ which avoids significant overlap with $\Span\set{\ket{000},\ket{111}}$ in $S$? The answer is given by the Traversal Lemma~\cite{gharibianGroundStateConnectivity2015,gossetQCMAHardnessGround2017}, which states that if two states $\ket{\psi}$ and $\ket{\phi}$ are ``orthogonal enough'', then any $2$-local evolution mapping from $\ket{\psi}$ to $\ket{\phi}$ \emph{must} leave $\Span(\ket{\psi},\ket{\phi})$; this is depicted in \Cref{fig:traversallemma}. Formally, the lemma says that for any pair of $k$-orthogonal\footnote{States $\ket{\psi}$ and $\ket{\phi}$ are $k$-orthogonal if they remain orthogonal under application of any single $k$-local unitary. }states $\ket{\psi}$ and $\ket{\phi}$, any sequence of $m$ $k$-local gates mapping $\ket{\psi}$ $\epsilon$-close to $\ket{\phi}$ must have overlap at least $(1-\epsilon)^2/m^2$ with the orthogonal complement of $\Span\set{\ket{\psi},\ket{\phi}}$. 

As an aside, the lower bound $(1-\epsilon)^2/m^2$ of the Traversal Lemma grows with the number of number of gates, $m$. This is not an accident --- in the case of \Cref{fig:traversallemma}, for example, it is tight~\cite{gharibianGroundStateConnectivity2015}. In fact, a more general statement holds --- for \emph{any} sufficiently Lipschitz continuous path on the hypersphere, one can follow said path within inverse exponential accuracy using $O(1)$-local gates, at the expense of blowing up the number of gates $m$ to also grow exponentially. One may interpret this in terms of quantum error-correcting codes as follows: One can  corrupt a ground state/codeword $\ket{\psi_1}$ of any (e.g.) stabilizer Hamiltonian $H$ via \emph{local} gates into a second codeword $\ket{\psi_2}$ \emph{without being detectable}, at the expense of making the local corruption process exponentially long.
\begin{open question}
  The Traversal Lemma is a general statement about the geometry of $k$-local computation paths, which \emph{a priori} has nothing to do with GSCON. Is there an application of the lemma outside the study of GSCON?
\end{open question}

\paragraph{Hardness of approximation.} As mentioned in \Cref{scn:QMA}, whether a quantum PCP theorem for LH holds remains a major open question. Classically, one of the major appeals of the PCP theorem~\cite{aroraProbabilisticCheckingProofs1998,aroraProofVerificationHardness1998} is to obtain hardness of approximation results for NP-complete problems. This raises the question: \emph{Can one already obtain hardness of approximation for a quantum complexity class without a corresponding ``quantum PCP theorem''?} The answer is yes, and there are two quantum classes known to fit this bill: $\cqs$~\cite{gharibianHardnessApproximationQuantum2012a} (a quantum analogue of $\Sigma_2^p$ with a classical $\exists$ proof, quantum $\forall$ proof, and quantum verifier), and QCMA~\cite{bittelOptimalDepthVariational2023}. We now discuss the QCMA-hard to approximate problem, which is not only natural, but particularly relevant in the current era of Noisy Intermediate Scale Quantum (NISQ) computation.\\

\begin{figure}[t]
    \begin{center}
      \includegraphics[height=4.5cm]{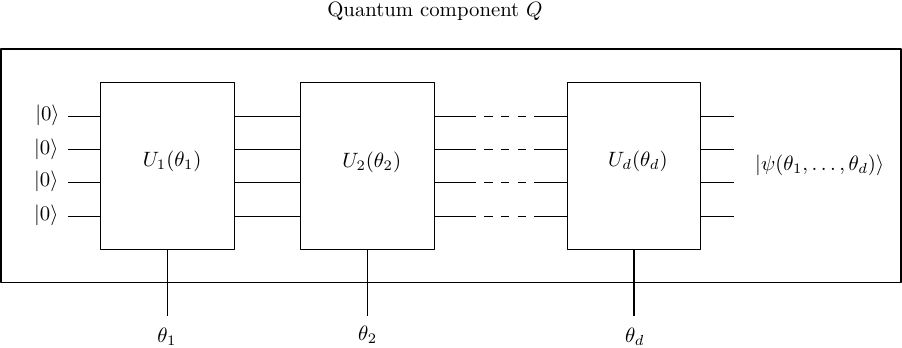}  
    \end{center}
    \caption{The heart of VQA, a parameterized circuit. Each box $U_i$ specifies a rotation axis, with $\theta_i$ the rotation angle. One typically fixes the ansatz $(U_1,\ldots, U_d)$, and subsequently attempts to optimize the angles $(\theta_1,\ldots, \theta_d)$.}
    \label{fig:VQA}  
  \end{figure}
\noindent\emph{Variational Quantum Algorithms (VQA).} Arguably the leading algorithmic framework in today's NISQ era is that of \emph{Variational Quantum Algorithms (VQAs)}~\cite{cerezoVariationalQuantumAlgorithms2021}. In a nutshell, VQAs are simply \emph{parameterized quantum circuits}, as depicted in \Cref{fig:VQA}. To explain, recall that a unitary operator is just a rotation in high-dimensional Hilbert space, i.e. is of form $U=e^{i\theta H}$, where Hamiltonian $H$ encodes the rotation axis, and $\theta$ the rotation angle. In a VQA, depending on the computational problem one wishes to solve, or the hardware platform at one's disposal, one first fixes the set of rotation \emph{axes} $H_1$ through $H_d$ --- this is called the \emph{ansatz}. The name of the game is now to find ``good'' rotation \emph{angles}, $\theta_1$ through $\theta_d$, so that the output state
\begin{equation}\label{eqn:vqastate}
  \ket{\psi(\theta_1,\ldots,\theta_d)}:=U_d(\theta_d)\cdots U_2(\theta_2) U_1(\theta_1)\ket{0\cdots 0}:=e^{iH_d\theta_d}\cdots e^{iH_2\theta_2}e^{iH_1\theta_1}\ket{0\cdots 0}
\end{equation}
produces ``good'' measurement results for the problem at hand. In practice, this setup is often run in a hybrid feedback loop~\cite{cerezoVariationalQuantumAlgorithms2021,chenComplexityNISQ2023}: The angles $\set{\theta_i}$ are chosen variationally (i.e. heuristically), the circuit is run on a quantum device and $\ket{\psi(\theta_1,\ldots,\theta_d)}$ is measured in the standard basis, and these results are fed back into the next round of variationally choosing new rotation angles. The dominance of VQA's in the NISQ era is arguably due primarily to two factors: (1) It is ``relatively easy''\footnote{Briefly, this stems from the fact that VQAs may be viewed as a (generalization of) truncated Trotterized quantum adiabatic algorithms~\cite{farhiQuantumApproximateOptimization2015}. Quantum adiabatic algorithms, in turn, are among the most natural quantum frameworks for attempting to solve classical optimization problems~\cite{farhiQuantumComputationAdiabatic2000}.} to \emph{set up} the framework in order to attempt to solve classical combinatorial optimization problems (though optimizing the angles is hard!~\cite{bittelTrainingVariationalQuantum2021a,bittelOptimalDepthVariational2023}), and (2) by its very design, it is amenable to low-depth constructions (i.e. one just restricts the number of boxes~\footnote{For clarity, by ``depth'' in VQA, one is not referring to the standard notion of circuit depth, but the parameter $d$ in \Cref{fig:VQA}} $U_i$), which are crucial for NISQ devices, which are prone to noise the deeper the circuit gets. The second of these is particularly relevant for our discussion here --- since depth is such an important bottleneck for NISQ, what is the complexity of finding the \emph{optimal} depth for a given ansatz?\\

\noindent\emph{Depth minimization for VQAs.} The answer is QCMA-hard, even to any reasonable relative error. To formalize this, the input is a set of rotation axes (local Hamiltonians) $\set{G_i}$, a local observable $M$, and depth thresholds $d_1<d_2$. The goal is to decide if there are at most $d_1$ angles $(\theta_1,\ldots, \theta_{d_1})\in \R^{d_1}$ and a corresponding sequence $(H_1,\ldots, H_{d_1})$ of Hamiltonians from $G$ (repetitions permitted) such that $\ket{\psi(\theta_1,\ldots,\theta_{d_1})}$ (\Cref{eqn:vqastate}) has ``good'' expectation against $M$, or if all such lists of at most $d_2$ angles and axes have ``bad'' expectation against $M$. It turns out this problem is QCMA-hard, even if $d_2/d_1\in\Omega(N^{1-\epsilon})$ for any fixed $\epsilon>0$, for $N$ the input size~\cite{bittelOptimalDepthVariational2023}. The proof begins with a clever idea of Umans~\cite{umansHardnessApproximatingSpl1999}, who used \emph{dispersers} to give hardness of approximation for $\Sigma_2^p$ \emph{without} resorting to a PCP. This can be ``quantized'' to give~\cite{gharibianHardnessApproximationQuantum2012a} QCMA-hardness of approximation for a seemingly artificial problem denoted QMSA: Given a quantum circuit $V$ accepting a non-empty monotone set of classical proofs, what is the minimum Hamming weight proof accepted by $V$? The key insight is now that Hamming weight minimization can be encoded into VQA depth minimization as follows. One defines two sets of VQA Hamiltonians: $P$ (Hamiltonians for ``setting proof bits''), and $Q$ (Hamiltonians for simulating gates from $V$). These are defined using ideas similar to Kitaev's circuit-to-Hamiltonian construction (\Cref{scn:QMA}) (except now multiple clocks are needed to obtain the desired hardness of approximation ratio). So, an example of a Hamiltonian in $P$ is
   \begin{align}
                P_j&\coloneqq X_{A_j}\otimes \ketbra{1}{1}_{C_{j}}\otimes \ketbra{1}{1}_{D_{\absD}}\label{eqn:flip}
   \end{align}
which roughly says: If clock 1 (register $C$) is at time $j$ \emph{and} clock 2 (register $D$) is at time $\abs{D}$ (more on clock 2 shortly), then flip the $j$th qubit of register $A$ via a Pauli $X$ gate. An example of a Hamiltonian in $Q$ is
                \begin{align}
                    Q_j &\coloneqq (V_j)_{AB}\otimes\ketbra{01}{10}_{C_{\absA+j,\absA+j+1}}+ (V_j^\dagger)_{AB}\otimes\ketbra{10}{01}_{C_{\absA+j,\absA+j+1}},
                \end{align}
which allows application of gate $V_j$ of $V$ to registers $AB$, while updating clock 1 from time $\absA+j$ to $\absA+j+1$. The point is that the minimum number of Hamiltonian evolutions needed from $P$ corresponds to the minimum Hamming weight of a proof $y$ accepted by $V$, and this was precisely the parameter in which QMSA gives us hardness of approximation. This sketch itself does not obtain the desired approximation ratio $N^{1-\epsilon}$, but gives the basic principle. 

\paragraph{Oracle separations between QMA and QCMA.} Finally, we have discussed the question ``\emph{what good is a classical proof to a quantum veriifer}'', and ultimately, this boils down to whether QCMA equals QMA. Separating these two would, unfortunately (?), also  separate P from PSPACE (since $\NP\subseteq \QCMA\subseteq\QMA\subseteq\PSPACE$). So, as all good complexity theorists, we consider \emph{oracle} separations. On the bright side, such oracle separations do exist~\cite{aaronsonQuantumClassicalProofs2007,feffermanQuantumVsClassical2018,natarajanDistributionTestingOracle2023} (see also \cite{bassirianPowerNonstandardQuantum2023}). On the less bright side, the ideal separation would utilize a \emph{classical} oracle, i.e. oracles as in Grover's algorithm~\cite{groverFastQuantumMechanical1996}, which implements a Boolean function $f:\set{0,1}^n\rightarrow \set{0,1}$. The separations~\cite{aaronsonQuantumClassicalProofs2007,feffermanQuantumVsClassical2018,natarajanDistributionTestingOracle2023}  have made significant progress towards this goal, using the following classes of oracles: (1) A unitary oracle, i.e. black-box $U$ mapping $\ket{x}\mapsto U\ket{x}$~\cite{aaronsonQuantumClassicalProofs2007}, (2) an in-place permutation oracle, i.e. black-box $U_\pi$ mapping $\ket{x}\mapsto\ket{\pi(x)}$ for $\pi$ a permutation~\cite{feffermanQuantumVsClassical2018}, and (3) \emph{distributions} over classical oracles~\cite{natarajanDistributionTestingOracle2023}. In the latter, the goal is to distinguish between two classes of \emph{distributions} $D_{\yes}$ and $D_{\no}$ over classical oracles. Given an oracle $O$ drawn from $D_{\yes}$, a QMA prover can send a quantum proof depending only on $D_{\yes}$ (and not the particular oracle, $O$), whereas a QCMA verifier cannot distinguish the two cases, \emph{assuming} (like for the QMA prover) its proof can only depend on the distribution, not $O$ itself.
\begin{open question}
  Is there a classical oracle separating QCMA from QMA?
\end{open question}

\section{Bashful: QMA with perfect completeness ($\QMAo$)}\label{scn:QMAo}

In \Cref{scn:QCMA}, we discussed how $\QCMA=\QCMA_1$, which begs the question: \emph{Does QMA also have perfect completeness, i.e. does $\QMA$ equal $\QMAo$?} The resilience of this question earns $\QMAo$ the title of \emph{Bashful}: That quiet student sitting in the back of the class, who at opportune moments, nails the answer to an important question, but otherwise prefers their privacy and comfort. Introduced by Bravyi~\cite{bravyiEfficientAlgorithmQuantum2006}, $\QMAo$ is defined exactly as one expects:
\begin{definition}[QMA with perfect completeness ($\QMAo$)]\label{def:QMAo}
  A promise problem $\A=(\ayes,\ano)$ is in QMA if there exists a P-uniform quantum circuit family $\set{V_n}$ and polynomials $p,q:\N\rightarrow\N$ satisfying the following properties. For any input $x\in\set{0,1}^n$, $V_n$ takes in $n+p(n)+q(n)$ qubits as input, consisting of the input $x$ on register $A$, $p(n)$ qubits initialized to a quantum proof $\ket{\psi}\in\Cm{p(n)}$ on register $B$, and $q(n)$ ancilla qubits initialized to $\ket{0}$ on register $C$. The first qubit of register $C$, denoted $C_1$, is the designated output qubit, a measurement of which in the standard basis after applying $V_n$ yields the following:
  \begin{itemize}
      \item (Completeness) If $x\in \ayes$, $\exists$ proof $\ket{\psi}\in\Cm{p(n)}$ that $V_n$ accepts with probability $1$.
      \item (Soundness) If $x\in \ano$, then $\forall$ proofs $\ket{\psi}\in\Cm{p(n)}$, $V_n$ accepts with probability $\leq 1/3$.
  \end{itemize}
\end{definition}
\noindent As a sanity check, the careful reader should inspect this definition, and (1) spot the single difference between QMA and $\QMAo$, and (2) realize that I lied when I said ``$\QMAo$ is defined exactly as one expects'', for the naive definition above is not quite correct. For in order to make any circuit-based complexity class well-defined, one needs to have a notion of a universal gate set, i.e. \emph{which} gates is the P-uniform quantum circuit family $\set{V_n}$ allowed to use? This is no problem for QMA and QCMA, since their promise gap and \emph{im}perfect completeness allow them to leverage standard universal gate sets based on the Solovay-Kitaev theorem~\cite{kitaevQuantumComputationsAlgorithms1997} (which necessarily make $\epsilon$-error in generating a dense subset of $SU(2)$). This raises the question:
\begin{open question}\label{q:qmaogates}
  Does $\QMAo$ have a universal gate set? 
\end{open question}
\begin{hint}
    If this question has a positive answer, it would presumably require a more ``computational'' approach (e.g.~\cite{aharonovSimpleProofThat2003})  than the Solovay-Kitaev theorem. In words, it is not reasonable to expect that there exists a finite fixed gate set which can simulate all of SU$(2)$ \emph{perfectly}. However, it seem plausible that there exists a finite fixed gate set $U$ so that: Given any $\QMAo$ verifier/optimal proof pair $(V,\ket{\psi})$ using some gate-set $G$, there is a poly-time Turing machine mapping $V$ to an entirely \emph{different} $\QMAo$ verifier/optimal proof pair $(V',\ket{\psi'})$ using gate set $U$, so that $V$ accepts $\ket{\psi}$ with certainty if and only if $V'$ accepts $\ket{\psi'}$ with certainty.
\end{hint}
\noindent Without a resolution to Question \ref{q:qmaogates}, we have to settle for a different version of $\QMAo$ for each possible gate set. A common choice, for example, is $\set{H,T,CNOT}$~\cite{gossetQuantum3SATQMA1Complete2013a}.

\paragraph{Complete problems.} Like QCMA, complete problems for $\QMAo$ are fewer and farther between than for QMA. However, I did say above that $\QMAo$ ``at opportune moments, nails the answer to an important question'', and in this regard I was referring to the Quantum SAT (QSAT) problem, which is $\QMAo$-complete~\cite{bravyiEfficientAlgorithmQuantum2006,gossetQuantum3SATQMA1Complete2013a}. (Most recently, there is also an interesting line of work showing that, of all places to find a connection, $\QMAo$ captures the complexity of determining homology groups of simplicial complexes (!)~\cite{cadeComplexitySupersymmetricSystems2021,crichignoCliqueHomologyQMA1hard2022}.) Intuitively, QSAT is to LH as classical SAT is to MAX-SAT --- given a set of quantum clauses $H_i$, is it possible to satisfy them ``perfectly''? Here, ``perfectly'' means the ground state $\ket{\psi}$ is a \emph{simultaneous} ground state of \emph{each} local term, i.e. for all $i$, $\bra{\psi}H_i\ket{\psi}=\lmin(H_i)$ (note the subscript $i$). Such Hamiltonians $H=\sum_i H_i$ are called \emph{frustration-free}. Equivalently, one defines QSAT by simplifying the definition of LH as follows: Each $H_i$ is now a \emph{projector} (i.e. Hermitian and eigenvalues in $\set{0,1}$), and the YES case now reads ``there exists $\ket{\psi}$ such that $\bra{\psi}H\ket{\psi}=0$''. 

It turns out the complexity of QSAT closely mirrors that of SAT in many respects. For example, analogous to the NP-completeness of $3$-SAT, $3$-QSAT is $\QMAo$-complete~\cite{gossetQuantum3SATQMA1Complete2013a}. Likewise, just as $2$-SAT is in P, so is $2$-QSAT~\cite{bravyiEfficientAlgorithmQuantum2006}, and it can in fact also be solved in linear time~\cite{aradLinearTimeAlgorithm2016a,beaudrapLinearTimeAlgorithm2016a}.
But wait --- how can the smallest eigenvalue of an exponentially large matrix be efficiently computed, if we \emph{cannot even write down} the corresponding eigenvector efficiently? It turns out that for $2$-QSAT (but not for $2$-LH!), this latter assumption is wrong --- we \emph{can} efficiently write down the ground state of a frustration-free $2$-local Hamiltonian~\cite{bravyiEfficientAlgorithmQuantum2006,chenNogoTheoremOneway2011,jiCompleteCharacterizationGroundspace2011}. At the heart of this turn of events is a simple classical observation (which also underlies the Aspvall-Plass-Tarjan linear-time algorithm for $2$-SAT~\cite{aspvallLineartimeAlgorithmTesting1979}): A $2$-SAT clause $x\vee y$ is equivalent to $(\overline{x}\implies y)\wedge(\overline{y}\implies x)$, for $\overline{x}$ the complement of bit $x$. Thus, assignments for $2$-SAT can be deterministically propagated\footnote{Contrast with $3$-SAT clause $x\vee y \vee z$, where if $x=0$, it is not clear whether we should set $y=1$ or $z=1$.} from $x$ to $y$ --- either $x=1$, in which no propagation is required (since the clause is already satisfied), or $x=0$, in which case we must set $y=1$. This allows one to embed the $2$-SAT instance $\phi$ into a directed graph, so that is unsatisfiable if and only if there is a cycle in which some variable is forced to be set to \emph{both} $0$ and $1$~\cite{aspvallLineartimeAlgorithmTesting1979}. Quantumly, we can generalize this trick to apply to \emph{tensor product} assignments $\ket{\psiprod}=\ket{\psi_1}\otimes\ket{\psi_2}\in\base\otimes\base$. Namely, to any rank-$1$ Quantum $2$-SAT clause $H_{12}=\ketbra{\psi}{\psi}$ with Schmidt decomposition\footnote{A Schmidt decomposition is just the Singular Value Decomposition in disguise, with $\set{s_0,s_1}$ being singular values of a certain reshuffled matrix corresponding to $\ket{\psi}$, $\set{\ket{a_0},\ket{a_1}}$ the set of left singular vectors, and $\set{\ket{b_0},\ket{b_1}}$ the right singular vectors.} $\ket{\psi}=s_0\ket{a_0}\ket{b_0}+s_1\ket{a_1}\ket{b_1}$, one associates the $2\times 2$ transfer matrix $T_\psi=s_1\ketbra{b_0}{a_1}+s_0\ketbra{b_1}{a_0}$\cite{bravyiEfficientAlgorithmQuantum2006,laumannc.r.PhaseTransitionsRandom2010}. Then, analogous to the classical setting, given assignment $\ket{\psi_1}$ onto qubit $1$, there is a \emph{unique}\footnote{If either Schmidt coefficient $s_0$ or $s_1$ equals $0$, which includes the classical $x\vee y$ clause, then again this propagation can fail to take place if $T\ket{\psi_1}=0$, meaning $H_{ij}$ is already satisfied by $\ket{\psi_1}$ alone.} assignment $\ket{\psi_2}=T_\psi\ket{\psi_1}$ so that $H_{12}\ket{\psi_1}\ket{\psi_2}=0$. One can then play a similar game involving analysis of cycles~\cite{laumannc.r.PhaseTransitionsRandom2010,beaudrapLinearTimeAlgorithm2016a} to solve the $2$-QSAT instance\footnote{For clarity, the ground state of a $2$-local frustration-free Hamiltonian $H$ is  in tensor product form, up to application of a set of $2$-local unitaries which can be efficiently computed given $H$~\cite{bravyiEfficientAlgorithmQuantum2006,aradLinearTimeAlgorithm2016a}.}, although \emph{how} the cycles are used is  different in the quantum setting.

Finally, let us complete our discussion of $\QMAo$-hardness of QSAT. In the case of LH, obtaining QMA-hardness for $2$-local constraints on \emph{qubits} typically requires perturbation theory, which while being tricky to apply, is at least nowadays standardized via the Schrieffer-Wolff transformation~\cite{bravyiSchriefferWolffTransformation2011}. Obtaining $\QMAo$-hardness of $3$-QSAT on qubits~\cite{gossetQuantum3SATQMA1Complete2013a}, on the other hand, remains a nasty ordeal, as the frustration-free requirement rules out the use of perturbation theory (which by design requires a frustrated ground space). Instead, one must manually design involved Hamiltonian gadgets which indirectly capture the logic of the $\QMAo$ verifier being embedded. As for $2$-QSAT, we have seen it is in P for the case of \emph{qubits}. Analogous to $2$-LH, however, it remains $\QMAo$-hard on the $1D$-chain on qu\emph{d}its of dimension $12$~\cite{nagajLocalHamiltoniansQuantum2008}. This leads to a final frontier --- what is the complexity of $2$-QSAT for local dimensions $2<d<12$? If we leave the 1D setting and consider general graphs, we do know that $2$-QSAT where each constraint acts on a qubit-qutrit pair, i.e. $(2,3)$-QSAT, is at least NP-hard~\cite{nagajLocalHamiltoniansQuantum2008}. 
\begin{open question}
  Is $(2,3)$-QSAT $\QMAo$-hard?
\end{open question}
\noindent In this direction, it is known that $(3,5)$-QSAT (i.e. all constraints on 3d and 5d particle pair) is $\QMAo$-hard~\cite{eldarQuantumSATQutritCinquit2008}. This has recently been improved to $\QMAo$-hardness of $(2,5)$-QSAT and $(3,4)$-QSAT~\cite{gharibianQuantumSATDimensional}.

\paragraph{Product states, QSAT, and TFNP.} As seen above, product state assignments have played an important role in the study of QSAT as harbingers of efficiently solvable cases~\cite{bravyiEfficientAlgorithmQuantum2006,laumannProductGenericRandom2010,laumannc.r.PhaseTransitionsRandom2010}. This trend continues --- for example, it was recently shown~\cite{montanaroTestingQuantumSatisfiability2023} that a ``property testing'' version of $k$-QSAT is solvable in BPP. In this model, one is promised that the given $k$-QSAT instance is either satisfiable, or is ``far from satisfiable'' by a product state (note the YES case places no requirement on product states). The poly-time algorithm is possible because, even in the YES case, one can prove that it suffices to randomly check for satisfiability by a product state on a randomly chosen $O(1)$-size subsystem (given the property-testing promise). So, QSAT with product state assignments is easy, right? Well, \emph{no}, for the obvious reason that one can easily embed a classical SAT formula $\phi$ into a QSAT instance $H$, so that $\phi$ is satisfiable if and only if $H$ has a null state which is a product state. Ok, but what if I promise you that \emph{there is} a satisfying product state $\ket{\psiprod}$ for a given $H$? By definition, this is in NP, but cannot be NP-hard, since a solution is guaranteed to exist. Surely, we can find $\ket{\psiprod}$ efficiently then? 

\emph{Not so fast.} Just like 1D $2$-QSAT on qudits remains $\QMAo$-hard in contrast to the poly-time solvability of $2$-SAT on a line, here we find another stark quantum departure from the classical SAT landscape, albeit down a different path. A textbook example of a $3$-SAT instance $\phi$ which is easy to solve is one with a \emph{System of Distinct Representatives (SDR)}. Formally, depict $\phi$ as a bipartite graph $G=(V,W,E)$, such that each $v\in V$ corresponds to a variable of $\phi$, and each $w\in W$ to a clause of $\phi$, and we place edge $(v,w)\in E$ if variable $v$ occurs in clause $w$. In this setting, an SDR means there is a matching of size $\abs{W}$, i.e. each clause $w$ has a unique variable matched to it. A $3$-SAT instance with an SDR is trivially solvable --- for any clause $c=(x\vee y\vee z)$ with matched variable (say) $x$, since $x$ is not matched to any other clause of an SDR, we can set $x=1$ to satisfy $c$. Moreover, the satisfying assignment can be efficiently computed, since the matching can be efficiently found when it exists via, e.g.,  reduction to network flow~\cite{jrMaximalFlowNetwork1956}. Ok, so now let's play the quantum version of this game --- $3$-QSAT instances with an SDR turn out to \emph{also} be always satisfiable by a product state~\cite{laumannProductGenericRandom2010}. So, can we find the satisfying product state efficiently? Partial positive~\cite{aldiEfficientlySolvableCases2021} and negative~\cite{goerdtMatchedInstancesQuantum2019} progress suggested the problem ``seems hard''. \\

\noindent \emph{Enter TFNP.} Classically, the rich theory of Total Function NP (TFNP)~\cite{megiddoTotalFunctionsExistence1991a} and its subclasses (PLS, PPA, PPP, PPAD~\cite{johnsonHowEasyLocal1988a,papadimitriouComplexityParityArgument1994}) is set up to deal with precisely this setting --- NP problems which are guaranteed to have a witness due to some mathematical principle\footnote{For example, Polynomial Pigeonhole Principle (PPP) is the set of search problems for which a solution is guaranteed to exist by the Pigeonhole Principle.}, but \emph{finding} the witness appears intractable. For example, finding Brouwer fixed points~\cite{papadimitriouComplexityParityArgument1994} and Nash equilibria~\cite{daskalakisComplexityComputingNash2006,chenSettlingComplexityComputing2009} are both famously PPAD-complete, and thus believed intractable, even though a fixed point and Nash equilibrium are always guaranteed to exist, respectively. It turns out~\cite{aldiQuantumComplexityTheory} that lurking behind QSAT with SDR is a ``new'' mathematical principle --- Bézout's theorem,  which in its original form from
1779, counts the number of common zeros of $n$ polynomials in $n$ variables~\cite{morganHomotopySolvingGeneral1987}. Specifically, the problem of finding a product state solution to \emph{any} QSAT instance (with or without an SDR), dubbed PRODSAT, can be written as a system of multihomogenous equations\footnote{A multihomogenous polynomial is one whose variable sets can be partitioned into sets $Z_j$, so that for any $Z_j$, viewing all variables not in $Z_j$ as constants yields a homogeneous polynomial.}. Then, the multihomogeneous extension~\cite{shafarevichBasicAlgebraicGeometry1974} of B\'{e}zout's theorem roughly says that a system of multihomogenous equations has at least $B$ solutions, for $B$ the \emph{B\'{e}zout number}. And one can show that in the homogeneous encoding of a PRODSAT system, $B$ equals none other than the number of (weighted) SDRs in the original QSAT instance. By defining a new complexity class in TFNP to capture this setting with $B>0$, dubbed Multi-Homogeneous Systems (MHS), one can show that QSAT with SDR is MHS-complete~\cite{aldiQuantumComplexityTheory}, and thus presumably intractable.
\begin{open question}
  What is the relationship between MHS and other subclasses of TFNP, such as PPAD or PPP?
\end{open question}

\section{Grumpy: QMA with unentangled provers ($\QMAt$)}\label{scn:QMAt}

Ah, $\QMAt$, the gift that keeps on giving. If there ever was a complexity class that told you to shove it each time you tried to study it, $\QMAt$ is surely it. Introduced by Kobayashi, Matsumoto, and Yamakami~\cite{kobayashiQuantumMerlinArthurProof2003}, $\QMAt$ is defined as $\QMA$, except where the proof is promised to be in tensor product form across a pre-specified cut $A$ versus $B$, i.e. $\ket{\psi}=\ket{\psi_1}_A\otimes\ket{\psi_2}_B$. Formally:

\begin{definition}[QMA with unentangled provers ($\QMAt$)]\label{def:QMAt}
  A promise problem $\A=(\ayes,\allowbreak\ano,\ainv)$ is in $\QMAt$ if there exists a P-uniform quantum circuit family $\set{V_n}$ and polynomials $p,q:\N\mapsto\N$ satisfying the following properties. For any input $x\in\set{0,1}^n$, $V_n$ takes in $n+2p(n)+q(n)$ qubits as input, consisting of the input $x$ on register $A$, quantum proof $\ket{\psi_1}_{A}\otimes \ket{\psi_2}_{B}\in(\Cm{p(n)})^{\otimes 2}$ on registers $A\otimes B$, and $q(n)$ ancilla qubits initialized to $\ket{0}$ on register $C$. The first qubit of register $C$, denoted $C_1$, is the designated output qubit, a measurement of which in the standard basis after applying $V_n$ yields the following:
  \begin{itemize}
      \item (Completeness) If $x\in \ayes$, $\exists$ proof $\ket{\psi_1}_{A}\otimes \ket{\psi_2}_{B}\in(\Cm{p(n)})^{\otimes 2}$, such that $V_n$ accepts with probability $\geq 2/3$.
      \item (Soundness) If $x\in \ano$, then $\forall$ proofs $\ket{\psi_1}_{A}\otimes \ket{\psi_2}_{B}\in(\Cm{p(n)})^{\otimes 2}$, $V_n$ accepts with probability $\leq 1/3$.
  \end{itemize}
\end{definition}
\noindent Not much is known about $\QMAt$, but what \emph{is} known is very intriguing (which is why the field keeps coming back to it, even though we know it's not good for us). \emph{Complete problems?} Maybe one or two? \emph{Upper bounds?} Essentially\footnote{Actually, there is a potentially stronger upper bound, $\QMAt\subseteq\QSigma_3\subseteq \NEXP$, for $\QSigma_3$ a quantum generalization of $\Sigma_3^p$~\cite{gharibianQuantumGeneralizationsPolynomial2018a}. $\QSigma_3$, however, is itself in need of more study.} just the trivial one, NEXP (yes, you read that right --- not even PSPACE is a known upper bound). Even establishing a basic property like weak error reduction took a lot of work~\cite{harrowEfficientTestProduct2010a}. So, what gives? Why is $\QMAt$ seemingly so much harder than $\QMA$?\\

\noindent\emph{Linear versus quadradtic optimization.} The acceptance probability of any quantum verification circuit $V$ given proof $\ket{\psi}$ can be written as $\trace(P\ketbra{\psi}{\psi})$, where $P\succeq 0$ is known as the \emph{POVM operator} encoding $V$. Formally, if $V$ acts on proof space $A$ (where $A$ can contain, e.g. quantum proofs like QMA, classical proofs like QCMA, or tensor product proofs like $\QMAt$), and has ancilla space $B$ initialized to all zeroes, one has
\begin{equation}\label{eqn:QMAt_POVM}
  P:= \left(I_A\otimes\bra{0\cdots 0}_B\right)V^\dagger\ketbra{1}{1}_{B_1}V \left(I_A\otimes\ket{0\cdots 0}_B\right)\in\lin{A}.
\end{equation}
In the case of a QMA verifier, the optimal acceptance probability over all proofs is thus
\begin{equation}
    \max_{\text{unit }\ket{\psi}}\trace(P\ketbra{\psi}{\psi})=\lmax(P).
\end{equation}
In words, QMA is  ``simply'' an eigenvalue problem for an exponentially large matrix, $P$, i.e. a \emph{linear} optimization. In contrast, the optimal acceptance probability for a $\QMAt$ verifier is
\begin{equation}\label{eqn:qmataccept}
  \max_{\text{unit }\ket{\psi_1}_A\otimes\ket{\psi_2}_B}\trace(P\ketbra{\psi_1}{\psi_1}_A\otimes \ketbra{\psi_2}{\psi_2}_B),
\end{equation}
which (due to the tensor product) is a \emph{quadratic} optimization over the exponentially large Hilbert space. And general optimizations of the form of \Cref{eqn:qmataccept} are known to be strongly NP-hard with respect go dimension~\cite{gurvitsClassicalDeterministicComplexity2003,ioannouComputationalComplexityQuantum2007,gharibianStrongNPhardnessQuantum2010}, unlike eigenvalue computation, which is poly-time in the dimension. \emph{No wonder $\QMAt$ is so grumpy.}

\paragraph{Complete problems and the Product Test.} Natural $\QMAt$-complete problems are exceedingly rare. Here are a few: (1) The \emph{Sparse Separable} Local Hamiltonian problem is $\QMAt$-complete~\cite{chaillouxComplexitySeparableHamiltonian2012a}. This is LH, except where the Hamiltonian is sparse\footnote{A \emph{sparse} Hamiltonian $H\in\herm{\Cn}$ is one where each row of $H$ has $\poly(n)$ non-zero entries, and given row index $i\in[2^n]$, one can efficiently compute all non-zero entries in row $i$.} (which is a natural condition), but the minimal energy optimization is not done over all ground states, but rather over all product states $\ket{\psi}=\ket{\psi_1}_A\ket{\psi_2}_B$ across a pre-specified cut (this is arguably less natural, but hey, we're talking about $\QMAt$ here). (2) Determining whether the output of an isometry\footnote{An isometry is like a unitary which is allowed to blow up the Hilbert space size, i.e. $U:\lin{A}\rightarrow\lin{A\otimes B}$ such that $UU^\dagger=I$.} is close in trace distance to a product state is $\QMAt$-complete~\cite{gutoskiQuantumInteractiveProofs2015a}. (3) Deciding whether there is a \emph{pure} global state $\ket{\psi}\in\Cn$ consistent with a set of local $k$-body density matrices is believed to be $\QMAt$-hard (it is certainly in $\QMAt$ since the SWAP test, which we discuss shortly, allows one to test purity~\cite{ekertDirectEstimationsLinear2002}). Note this is the pure state version of the QMA-complete Consistency problem~\cite{liuConsistencyLocalDensity2006}.

\begin{figure}[t]
  \begin{center}
    \hspace{7mm}
    \begin{minipage}[c]{0.45\textwidth}
    \[ \Qcircuit @C=1.5em @R=0.5em {
       \lstick{\ket{0}}    &\gate{H} & \ctrl{1}                       & \gate{H} &\meter\\
       \lstick{\ket{\psi}} &\qw      & \multigate{1}{\textup{SWAP}} & \qw      &\qw\\
       \lstick{\ket{\phi}} &\qw      & \ghost{\textup{SWAP}}     &\qw       &\qw\\
    }\]
    \end{minipage}  
    \hspace{5mm}
    \begin{minipage}[c]{0.45\textwidth}
      \includegraphics[width=5cm]{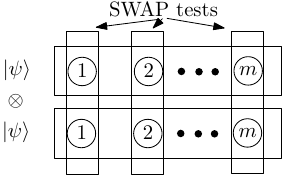}
    \end{minipage}
  \end{center}
    \caption{Left: The SWAP test. Right: Given two copies of an $m$-partite state $\ket{\psi}$, the Product test performs a SWAP test between the pair of $i$th systems from both copies, for all $i\in[m]$.}
    \label{fig:ProductStatetest}
  \end{figure}

A crucial ingredient for $\QMAt$-completeness of Sparse Separable LH is the Product test, which is also the key tool for proving weak error reduction for $\QMAt$, and that $\QMA(\poly)=\QMAt$~\cite{harrowEfficientTestProduct2010a} (i.e. $2$ Merlins are as powerful as $\poly(n)$ Merlins). For this, we first need the SWAP test (\Cref{fig:ProductStatetest}, left), which given physical copies of states $\ket{\psi}$ and $\ket{\phi}$, checks if $\ket{\psi}\approx\ket{\phi}$. Formally, the circuit outputs\footnote{If one instead inputs mixed states $\rho\otimes\sigma$, the test more generally outputs $0$ with probability $(1+\trace(\rho\sigma))/2$~\cite{ekertDirectEstimationsLinear2002}. Finally, if a \emph{non-tensor product} input is given, upon measuring $0$ ($1$) the circuit projects the input register onto the symmetric (anti-symmetric) subspace.} $0$ with probability $(1+\abs{\braket{\psi}{\phi}}^2)/2$. Moving to the Product test, assume we wish to simulate $\QMA(\poly)$ with $\QMAt$, i.e. we wish to simulate proof $\ket{v}:=\ket{\psi_1}\otimes\cdots\otimes\ket{\psi_m}$ with some $\ket{w}:=\ket{\phi}\otimes\ket{\phi}$. The difficulty is that in the NO case, a cheating prover may send a $\ket{v}$ which is \emph{not} in tensor product across the $m$ subsystems, breaking the soundness analysis for $\QMA(\poly)$. The Product test catches this, by asking the $\QMAt$ prover to send $\ket{w}=\ket{v}\otimes\ket{v}$, and performing ``column-wise'' SWAP tests as in \Cref{fig:ProductStatetest}. The intuition is simple --- for an honest prover, each ``column'' $i$ contains $\ket{\psi_i}\otimes\ket{\psi_i}$, so the SWAP test accepts with certainty. The original proof of soundness for the Product test was involved~\cite{harrowEfficientTestProduct2010a}, but there is now a simple, slick proof~\cite{soleimanifarTestingMatrixProduct2022a}.

\paragraph{Magic: Compressing classical proofs.} A surprising power of $\QMAt$ is its ability to ``compress'' (e.g.) NP proofs. For example, suppose we wish to verify that a graph $G=(V,E)$ is $3$-colorable. An NP prover would send the list of color assignments $c(v)$ to each vertex $v\in V$, which has $O(n)$ size. Quantumly, however, we could try to encode the colors \emph{in superposition}, i.e. via proof $\ket{\psi}=\frac{1}{\sqrt{n}}\sum_{v\in V}\ket{v}\ket{c(v)}\in\Cm{\log (n)}$, which only requires $\log (n)$ qubits. \emph{Can a QMA verifier verify this?} The answer is that no one knows, but what we \emph{do know} is that given \emph{two} copies $\ket{\psi}\ket{\psi}\in\Cm{2\log n}$, the answer is yes$^{\ast}$~\cite{blierQuantumCharacterizationNP2012,gallQMAProtocolsTwo2012}. (This crucially uses the SWAP test. Also, the $\QMAt$ proof system also requires only $O(\log n)$ ancilla space~\cite{blierQuantumCharacterizationNP2012}, a fact we will use shortly.) The catch$^{\ast}$ is that with such a short proof, we only know how to attain a $1/\poly(n)$ promise gap. This may seem like a footnote, but actually it will be a royal pain when it comes to bounding the power of $\QMAt$. For the only way we know how to improve the gap to a constant requires blowing up the proof size, e.g. with $O(\sqrt{n})$ total proof length a constant gap can be achieved~\cite{aaronsonPowerUnentanglement2009,chiesaImprovedSoundnessQMA2013}. But we also know that $\QMAt$ with $1/\exp(n)$ gap equals NEXP, obtained by blowing up the $3$-COLORING protocol~\cite{blierQuantumCharacterizationNP2012} to SUCCINCT $3$-COLORING. So, intuitively, \emph{if} we could have obtained a constant gap for NP with a $\log (n)$-size proof, we could presumably blown this up to a $\QMAt$ protocol for NEXP, obtaining $\QMAt=\NEXP$. \emph{Sigh.}\\

\noindent \emph{The essence of $\QMAt$?} A goal of this article was to distill the essence of each variant of quantum NP, and so far, the common ingredient we have seen re-used above is the SWAP test --- from weak error reduction to purity testing to compressing NP proofs. So, is ``$\QMAt=\text{ability to do SWAP tests}$''? Not quite --- at least for compressing NP proofs to length $O(\sqrt{n})$, the SWAP test can be avoided~\cite{chenShortMultiProverQuantum2010}! So, let us go back to the definition of $\QMAt$: By \Cref{eqn:qmataccept}, the power of $\QMAt$ stems from its ability to ``do quadratic things''. This, it turns out, can be leveraged by our good friend, a circuit-to-Hamiltonian construction.

To discuss this in the context of proof compression, we need a brief detour into \emph{streamed proofs}~\cite{gharibianQuantumSpaceGround2023}. Consider classical proof $y\in \set{0,1}^N$, which is streamed bit by bit to a quantum verifier $V$ acting on $O(\log N)$ qubits (but with size at least $\Omega(N)$ in order to read the whole proof). To formalize this, we imagine that in addition to its usual ancilla, the verifier has a designated $1$-qubit ``message register'' $M$ initialized to $\ket{0}$. At certain pre-determined time steps, the streaming prover may apply either Pauli $I$ or $X$ on $M$ in order to simulate preparation of the next streamed bit, $0$ or $1$, respectively. The verifier can then copy $M$ via CNOT into its private ancilla, if desired (no other joint operation with $M$ is allowed), and continue processing its full ancilla space (other than $M$). The key is that we may view the streaming procedure as a quantum circuit $W$ of length $O(N)$, acting on $\log(N)$ qubits, but in which the sequence of gates on $M$ is \emph{a priori unknown} to the verifier.

So, can we now apply Kitaev's circuit-to-Hamiltonian construction to $W$? Well, \emph{no}, because the construction requires advance knowledge of \emph{all} gates in $W$ when building the propagation Hamiltonian, $\hprop$, which we lack in the streaming setting. It's almost as if we are trying to build a \emph{history state} $\ket{\psihist}$ which has to predict the \emph{future}. While this seems paradoxical, it turns out that, by ``doing quadratic things'', a $\QMAt$ verifier can indeed allow a history state to ``encode the future''~\cite{gharibianQuantumSpaceGround2023}, as we next describe. \\

\noindent\emph{A quadratic circuit-to-Hamiltonian construction for $\QMAt$.} The kernel of the idea is best seen by considering its classical analogue. Suppose I, as verifier, wish to allow you, as prover, to stream either bit $0$ or $1$ to me, but I wish to prescribe your action via Boolean constraint \emph{ahead of time}. With only \emph{one} copy of your bit, $x\in\set{0,1}$, this is impossible: There are only two possible constraints I could put on your bit, $f(x)=x$ or $f(x)=\overline{x}$. The first forces you to stream $x=1$, and the second $x=0$; in both cases, you have no choice regarding the value of $x$. But if you send me \emph{two} copies $x,y\in\set{0,1}$ of your bit, I can play some old-school magic --- I can encode the EQUALS function $f(x,y)=(x\vee\overline{y})\wedge(\overline{x}\vee y)$. Now, you can encode \emph{either} $x=0$ or $x=1$, so long as $y=x$. Thus, you have freedom of choice for $x$, and we can embed this into Kitaev's construction as follows. 

The prover embeds the $O(N)$-gate long circuit $W$ into a $O(\log(N))$-qubit history state $\ket{\psihist}$ (this requires the clock to be encoded in binary, which gives a sparse Hamiltonian), except in order to ``quantumly simulate'' the EQUALS gadget above, the prover sends $\ket{\psihist}_A\otimes\ket{\psihist}_B$. Then, for each time step $t$ in which you streamed a bit, we add to Kitaev's $\hprop$ Hamiltonian (\Cref{eqn:hprop}) the gadget:
\begin{equation}\label{eqn:gad}
  (\HttI)_A\otimes (\HttiX)_B  +(\HttiX)_B\otimes (\HttI)_A \quad\text {for} \quad\HttV:=-V_{t+1}\otimes\ketbra{t+1}{t} - V_{t+1}^\dagger\otimes\ketbra{t}{t+1}.
\end{equation}
In words, this constraint is the quantum analogue of the (arithmetized) classical EQUALS function, $f(x,y)=x(1-y)+(1-x)y$. It is annihilated if and only if in time step $t$, the prover applies the \emph{same} gate (either $I$ or $X$) to \emph{both} copies of $\ket{\psihist}$, but which of $I$ or $X$ is actually applied is otherwise free. Thus, we have used a \emph{quadratic} gadget (due to the tensor product) to verify a $\QMAt$ proof, leveraging the very definition of $\QMAt$.

Since this ``quadratic circuit-to-Hamiltonian construction'' is agnostic to the choice of $N$, it can be scaled up to recover the NP~\cite{blierQuantumCharacterizationNP2012} ($N\in\poly(n)$) and NEXP~\cite{pereszlenyiMultiProverQuantumMerlinArthur2012} ($N\in\exp(n)$) compressed $\QMAt$ proof systems. However, while the Hamiltonian produced correctly acts on only $\log(N)$ qubits,  its \emph{promise} gap scales as $1/\poly(N)$. Thus, it cannot resolve the $\QMAt$ versus $\NEXP$ question, since we obtain an inverse exponential gap (relative to input size, $n$) in the NEXP case, i.e. one recovers\footnote{$\PreciseQMAt$ is $\QMAt$ with exponentially small gap.} $\PreciseQMAt=\NEXP$~\cite{pereszlenyiMultiProverQuantumMerlinArthur2012}. 
\begin{open question}
  Is $\QMAt=\NEXP$? Can other ``quadratic'' circuit-to-Hamiltonian constructions be developed to make progress\footnote{Aside: The most recent progress on $\QMAt$ versus $\NEXP$ is that if one forces the $\QMAt$ proof to have only \emph{non-negative} amplitudes, denoted $\QMA^+(2)$, then $\QMA^+(2)=\NEXP$~\cite{jeronimoPowerUnentangledQuantum2023}. However, this restriction is quite strong, as even $\QMA^+=\NEXP$~\cite{bassirianQuantumMerlinArthurProofs2023}.} towards this question?
\end{open question}

\section{Dopey: Stoquastic Merlin-Arthur (StoqMA)}\label{scn:StoqMA}

Stoquastic Merlin-Arthur is defined roughly as QMA, except the ``core'' of the verification circuit must be \emph{classical}, and instead, one is allowed to apply a single layer of parallel Hadamard gates in the first and last time steps. Frankly, the name ``Dopey'' chose itself. Introduced by Bravyi, DiVincenzo, Oliveira, and Terhal~\cite{bravyiComplexityStoquasticLocal2008}, it is formally defined as:

\begin{definition}[Stoquastic MA (StoqMA($\alpha$,$\beta$)]\label{def:stoqMA}
  A promise problem $\A=(\ayes,\ano)$ is in StoqMA if there exists a P-uniform quantum circuit family $\set{V_n}$ and polynomials $p,q:\N\rightarrow\N$ satisfying the following properties. For any input $x\in\set{0,1}^n$, $V_n$ takes in $n+p(n)+q(n)$ qubits as input, consisting of the input $x$ on register $A$, $p(n)$ qubits initialized to a quantum proof $\ket{\psi}\in\Cm{p(n)}$ on register $B$, and $q(n)$ ancilla qubits initialized to $\ket{0}$ on register $C$. The circuit $V_n$ has the following structure, where we slice the circuit into $L_n$ layers, with each layer consisting of a set of $2$-qubit gates which can be applied in parallel: The first layer consists only of parallel Hadamard gates applied on any desired subset of ancilla qubits. Layers $2$ to $L_n-1$ consist of classical gates, i.e. $X$, CNOT and Toffoli. Layer $L_v$ is parallel Hadamard gates applied on any desired subset of qubits. Measuring the designated output qubit, $C_1$, in the standard basis after applying $V_n$ yields the following:
  \begin{itemize}
      \item (Completeness) If $x\in \ayes$, $\exists$ proof $\ket{\psi}\in\Cm{p(n)}$ that $V_n$ accepts with probability $\geq\alpha$.
      \item (Soundness) If $x\in \ano$, then $\forall$ proofs $\ket{\psi}\in\Cm{p(n)}$, $V_n$ accepts with probability $\leq \beta$.
  \end{itemize}
  Above, $\beta-\alpha\geq 1/\poly(n)$.
\end{definition}
\noindent Note this is the only ``quantum NP'' we have defined for which completeness and soundness is \emph{not} arbitrary, i.e. we require thresholds $\alpha$ and $\beta$. This is because error reduction is not known to hold\footnote{The problem is the usual parallel verification fails, because no classical post-processing (including a majority vote) is allowed after layer $L_n$ of Hadamard gates.} for StoqMA; in fact, if one could reduce the error to $\alpha=1-o(1/\poly(n))$ versus $\beta=1-1/\poly(n)$, then $\StoqMA=\MA$~\cite{aharonovStoqMAVsMA2021} (see also~\cite{aharonovStoquasticPCPVs2019}).

\paragraph{The complexity of StoqMA.} Despite its awkward definition, StoqMA has burrowed itself deep into the quantum complexity landscape. Remember the ``quad''-chotomy theorem (\Cref{scn:QMA}) for LH? It involved three notions of hardness: NP-complete, StoqMA-complete, and QMA-complete. The corresponding canonical StoqMA-complete LH problem is known as the \emph{transverse field Ising model (TIM)},
\begin{equation}
  H=\sum_{(i,j)\in E} p_{ij} Z_i Z_j +\sum_k q_k X_k
\end{equation} 
where $p_{ij}$ and $q_k$ are polynomial weights\footnote{Formally, the weights $p_{ij}$ must be negative for this to qualify as a stoquastic Hamiltonian, but this can be achieved without loss of generality via conjugation by local $Z$ gates.} needed for hardness due to the use of perturbation theory~\cite{bravyiComplexityQuantumIsing2017}. (Contrast this with the NP-complete Max-Cut/Ising interaction $ZZ$, or the QMA-complete Quantum Max Cut interaction, $XX+YY+ZZ$.) At first glance, StoqMA also ``sits'' structurally in the right place, i.e.\footnote{The first containment holds because the MA verifier can simply not apply Hadamards in layer 1, and copy its output to $C_1$ before the Hadamard in layer $L_n$.} $\MA\subseteq\StoqMA\subseteq\QMA$. All seems good until we hit a serious speed bump --- StoqMA fails arguably the most basic requirement of ``quantum NP'', which is that it should contain quantum P, i.e. BQP. And yet, while BQP is not believed to be inside $\PH$ (indeed, there is an oracle separation~\cite{razOracleSeparationBQP2019}), StoqMA satisfies\footnote{Actually, $\StoqMA\subseteq\SBP\subseteq \AM$~\cite{bravyiMerlinArthurGamesStoquastic2006a}, for SBP the class Small Bounded-Error Probability.} $\StoqMA\subseteq\AM\subseteq\PH$~\cite{bravyiComplexityStoquasticLocal2008}.

\paragraph{Why does StoqMA seem ``easier'' than BQP?} The ``simplest'' resolution to this conundrum is to \emph{conjecture} that $\StoqMA=\MA$~\cite{aharonovStoqMAVsMA2021} (which is, frankly, plausible). But even if we don't go to this extreme, there is good intuition here, again via Hamiltonian complexity. Namely, the \emph{full} class of Hamiltonians which are StoqMA-complete~\cite{bravyiMerlinArthurGamesStoquastic2006a} (of which TIM is a special case) are those with real entries and all off-diagonal elements being non-positive\footnote{This is with respect to the standard basis. Deciding if a given Hamiltonian $H$ can be converted to stoquastic form under unitary conjugation (which preserves the spectrum of $H$) can be NP-hard~\cite{marvianComputationalComplexityCuring2019,klassenHardnessEaseCuring2020}. }. What this buys us is that the corresponding \emph{Gibbs density matrix} $\rho=e^{-\beta H}/\trace(e^{-\beta H})$ has \emph{non-negative} matrix elements. Roughly, the Gibbs state is a ``noisy/high-temperature'' analogue of the ground state $\ket{\psi}$, i.e. it is the state of the system $H$ in thermal equilibrium at temperature $T\propto 1/\beta$. Since the ground state $\ket{\psi}$ can be obtained by taking the limit $\beta\rightarrow \infty$ (i.e. temperature $T\rightarrow 0$), we obtain a strong property for $\ket{\psi}=\sum_x\alpha_x\ket{x}$ --- its amplitudes satisfy $\alpha_x\geq 0$ for all $x\in\set{0,1}^n$. In words, in a stoquastic system, $\ket{\psi}$ encodes a \emph{classical distribution}. \emph{Does this buy us anything?} Yes, in two respects. First,  Stoquastic $k$-QSAT, i.e. with a \emph{stoquastic} Hamiltonians $H$, is MA-complete~\cite{bravyiComplexityStoquasticFrustrationFree2010} for $k\geq 6$. This is in constrast to Stoquastic LH, which is $\StoqMA$-complete, and general QSAT, which is $\QMAo$-complete. Second, this ground state property of stoquastic Hamiltonians actually allows us to classically efficiently simulate a \emph{subset} of BQP. This requires a brief detour into adiabatic quantum computation.\\

\noindent\emph{Adiabatic quantum computation for stoquastic Hamiltonians: The frustration-free case.} The quantum adiabatic algorithm is a \emph{universal} model of quantum computation~\cite{aharonovAdiabaticQuantumComputation2004}, and works roughly as follows. The input is a pair of (say) local Hamiltonians $(\Hinit,\Hfinal)$, where the ground state $\ket{\psiinit}$ of $\Hinit$ can be prepared efficiently, and the ground state $\ket{\psifinal}$ of $\Hfinal$ encodes the solution to a computational problem we wish to solve. The standard approach to adiabatic computing now defines a linearly parameterized Hamiltonian $H(t)=(1-t)\Hinit + t\Hfinal$ for $t\in[0,1]$. Observe first that by definition of the word ``eigenvector'', $e^{iH(0)}\ket{\psiinit}\propto\ket{\psiinit}$, i.e. nothing happens. But if we slowly increase $t$ from $0$ to $1$ during this evolution (i.e. we evolve according to time-\emph{dependent} Hamiltonian $H(t)$), then $\ket{\psiinit}$ \emph{will} change. Indeed, the adiabatic theorem says that if this rate of change of $t$ scales as $1/\poly(\Delta)$, for $\Delta$ the minimum spectral gap of $H(t)$ over $t\in [0,1]$, then upon reaching $t=1$ we will have prepared $\ket{\psifinal}$. Since this model is universal, any BQP algorithm can be captured this way. 

Returning to the stoquastic setting, if local Hamiltonian $H=\sum_i H_i$ is \emph{both} stoquastic \emph{and} frustration-free, then the adiabatic algorithm can be efficiently simulated classically~\cite{bravyiComplexityStoquasticFrustrationFree2010}. Specifically, there is a classical algorithm which, for any $\delta>0$, can sample from a distribution which is total variation distance at most $\delta$ from the distribution obtained from measuring $\ket{\psifinal}$ in the standard basis. How does this algorithm leverage the fact that $\alpha_i\geq 0$ for $H$'s ground state, $\ket{\psi}$? The basic premise is that we can define a random walk on the \emph{support} $S$ of $\ket{\psi}$, i.e. all $x\in\set{0,1}$ such that $\alpha_x>0$, so that the transition probabilities $\pxy$ for moving from $x$ to $y$ can be classically computed. Formally, one defines 
\begin{equation}
    \pxy = \frac{\alpha_y}{\alpha_x}\bra{y}(I-\beta H)\ket{x} \text{ for }x\in S,
\end{equation}
where $\beta>0$ is chosen small enough so that $G$ is entry-wise non-negative and $\pxy\geq 0$. One can show that the stationary distribution of this walk is exactly what we want, $p_x=\alpha_x^2$. Finally, how does one actually compute $\alpha_x$ to begin with? It turns out that if\footnote{Note that this condition must hold for some $i$ if $\pxy>0$.} $\bra{y}H_i\ket{x}<0$ for one of the local stoquastic terms $H_i$, then 
\begin{equation}
  \frac{\alpha_y}{\alpha_x}=\sqrt{\frac{\bra{y}\Pi_i\ket{y}}{\bra{x}\Pi_i\ket{x}}},
\end{equation}
for $\Pi_i$ the projector onto the null space of $H_i$. But in a given walk step, we know $x$, and since $H_i$ is $O(1)$-local, there are only $\poly(n)$ many strings $y$ we need to check for the condition $\bra{y}H_i\ket{x}<0$. Thus, we can simulate the next step in the random walk from any $x$, \emph{even though we don't know $\alpha_x$ and $\alpha_y$}!

\paragraph{So, is StoqMA ``easy'' or ``hard''? Adiabatic evolution in the frustrated case.} The fact that stoquastic Hamiltonians have a ground state with non-negative amplitudes turned out to be crucial in simulating such Hamiltonians, but that was in the \emph{frustration-free} case. By now, the reader has probably noticed that the true theme of this article is ``don't get too comfortable studying quantum NP'',  and so here is your final twist: There is a superpolynomial oracle separation between the power of classical computation and adiabatic quantum computation with \emph{frustrated}, \emph{sparse} (as opposed to local) stoquastic Hamiltonians~\cite{hastingsPowerAdiabaticQuantum2021,gilyenSubExponentialAdvantage2021}. Roughly, this result has two main ingredients: A modified version of the ``welded-trees'' graph problem~\cite{childsExponentialAlgorithmicSpeedup2003} (which a quantum algorithm can solve efficiently given oracle access to the graph) and ``graph decorations''~\cite{hastingsPowerAdiabaticQuantum2021} (which complicate the graph sufficiently so that a classical algorithm cannot solve the problem). Note the oracle model ties in nicely with the sparse matrix setting, since recall we define the latter as being able to compute all $\poly(n)$ entries in any given row $i$ of the matrix efficiently. 

\begin{open question}
  Can one show a similar separation in power between classical computation and \emph{local} stoquastic Hamiltonians~\cite{gilyenSubExponentialAdvantage2021}? Note the latter setting would no longer necessitate an oracle, since the non-zero entries of local Hamiltonians can efficiently be computed by a classical computer. Of course, this also makes proving a lower bound harder, since one is no longer in the oracular setting.
\end{open question}

\section{Sneezy: Non-deterministic quantum poly-time (NQP)}\label{scn:NQP}

Finally, we have NQP. NQP is Sneezy, because$\ldots$ well, it came and went like a sneeze. Introduced by Adleman, DeMarrais, and Huang~\cite{adlemanQuantumComputability1997}, NQP is defined as follows:
\begin{definition}[Non-deterministic quantum polynomial time (NQP)~\cite{fennerDeterminingAcceptancePossibility1999}]\label{def:NQP}
  A language $L$ is in NQP if and only if there is a quantum Turing machine $V$ and a polynomial $p$ such that 
  \begin{equation}
      x\in L \iff \textup{Pr}[V \text{ accepts }x \text{ in }p(\abs{x}) \text{ steps}] \neq 0.
  \end{equation}
\end{definition}
\noindent The intuition above is that NP can be defined as the set of decision problems solvable by a randomized Turing machine $M$ which picks a uniformly random proof $y$, which it then verifies. Thus, in a YES case, $M$ accepts with probability at least $1/\exp(n)$, and in the NO case, with probability $0$. While this works for NP, things have played out differently for NQP. Soon after its definition, it was shown that $\NQP=\coCeP$~\cite{fennerDeterminingAcceptancePossibility1999}. Here, $\coCeP$ is the complement of $\CeP$, the counting class defined as follows: Language $L\in\CeP$ if there is a $\GapP$ function such that for any input $x$, $x\in L$ if and only if $f(x)=0$. Recall here that a $\GapP$ function computes the \emph{difference} between the number of accepting and rejecting paths of a non-deterministic Turing machine. To put this in context, $\CeP$ is hard for PH under randomized reductions~\cite{todaCountingClassesAre1992,taruiProbabilisticPolynomialsAC01993}. Thus, deciding an instance of an NQP language is also PH-hard under randomized reductions~\cite{fennerDeterminingAcceptancePossibility1999}. Finally, note NQP is not well-motivated from a \emph{physical} standpoint, in the sense that sampling access to the output of a quantum computing device only allows us to efficiently compute acceptance probabilities to within additive \emph{inverse polynomial} error.

\section*{Acknowledgements} I thank Anurag Anshu, Alex Grilo and Chinmay Nirkhe for helpful discussions. I am grateful for support from
the DFG under grant numbers 432788384 and 450041824, BMBF within the funding program ``Quantum Technologies - from Basic Research to Market'' via project PhoQuant (grant
number 13N16103), and project ``PhoQC'' from the programme “Profilbildung 2020”, an initiative of the Ministry of Culture and Science of the State of North Rhine-Westphalia.

\bibliographystyle{alpha}
\bibliography{Sev.bib}

\end{document}